\documentclass{elsarticle}

\usepackage{hyperref}  

\journal{Astroparticle Physics}

\usepackage{subcaption}
\usepackage{subcaption}
\usepackage{graphicx,placeins}
\usepackage{amsmath,amssymb}
\usepackage[multiple]{footmisc}

\begin{document}
	
\begin{frontmatter}
		
\title{Making light of gravitational-waves}
		
		\author{Justine Tarrant\corref{mycorrespondingauthor}}
		\cortext[mycorrespondingauthor]{Corresponding author}
		\ead{justine.tarrant@wits.ac.za}
		
		\author{Geoff Beck\corref{gb}}
		\ead{geoffrey.beck@wits.ac.za}
		
		\author{Sergio Colafrancesco\corref{sc}\footnote{This paper is dedicated to the memory of Prof. Sergio Colafrancesco.}}

		\address{School of Physics, University of the Witwatersrand,\\
			Private Bag 3, 2050, Johannesburg, South Africa}

		\begin{abstract}
			Mixing between photons and low-mass bosons is well considered in the literature. The particular case of interest here is with hypothetical gravitons, as we are concerned with the direct conversion of gravitons into photons in the presence of an external magnetic field. We examine whether such a process could produce direct low-frequency radio counterparts to gravitational-wave events. Our work differs from previous work in the literature in that we use the results of numerical simulations to demonstrate that, although a single such event may be undetectable without at least $10^5$ dipoles, an unresolved gravitational wave background from neutron star mergers could be potentially detectable with a lunar telescope composed of $10^3$ elements. This is provided the gravitational wave spectrum only experiences exponential damping above 80 kHz, a full order of magnitude above the limit achieved by present simulation results. In addition, the extrapolation cannot have a power-law slope $\lesssim -2$ (for 100 hours of observation time) and background and foregrounds must be effectively subtracted to obtain the signal. This does not make detection impossible, but suggests it may be unlikely. Furthermore, assuming a potentially detectable spectral scenario we show that, for the case when no detection is made by a lunar array, a lower bound, competitive with those from Lorentz-invariance violation, may be placed on the energy-scale of quantum gravitational effects. The SKA is shown to have very limited prospects for the detection of either a single merger or a background. 
		\end{abstract}
		
		
	\end{frontmatter}
	

	
	\section{Introduction}
	\label{sec:intro}
	Electromagnetic counterparts to gravitational-waves were well considered in the literature \cite{Schnittman:2014jpa,Essick:2014wwa,Pannarale:2014rea,Williamson:2014wma,Kasen:2014toa,Kamble:2014zya} prior to the first detection of compact mergers by the Laser Interferometer Gravitational-wave Observatory (LIGO), made in September 2015, and produced by the merger of two black holes. Furthermore, this correlation was considered by \cite{Dolgov2017,PhysRevD.99.044022} following these events. It wasn't until the discovery of counterparts associated with the fifth LIGO event \cite{TheLIGOScientific:2017qsa} detected in August 2017, that the field was revived as scientists poured over the new data obtained. The successful detection was only possible due to an immense, well coordinated, global collaboration. The GW170817 event was the first detection of a binary neutron star merger, and was long hoped for since such events were expected to form a kilonova and collimated outflows, called jets, which result in electromagnetic counterparts. 
	
	Emissions across the entire electromagnetic spectrum were detected \cite{GBM:2017lvd}, thus heralding the dawn of multi-messenger astronomy. This includes the observation of radio emissions in the broad band afterglow occurring when the gamma-ray burst interacts with the interstellar medium \cite{Alexander:2017aly}. In this work, we approach this problem from a different angle using the idea that gravitational-waves can themselves directly generate plausibly detectable low-frequency radio counterparts while propagating from their source to Earth.
	
	The idea that gravitational degrees of freedom may be converted into electromagnetic degrees of freedom is not a new one, see for example: \cite{Raffelt:1987im,Denisov1978,Gerlach:1974zz,Gretarsson_2018} and \cite{Gertsenshtein:1962}. Interest in this subject resulted in the need to find an indirect means of measuring gravitational-waves and also arose out of studies considering axion-photon mixing \cite{Peccei:1977hh,Kaplan:1985dv}. For example, early studies involved the scattering of electromagnetic fields off time-dependent gravitational fields, showing that there was a possible coherent interaction between linear gravitational-waves and electromagnetic waves in which energy could be transformed from one degree of freedom to the other \cite{Marklund:1999sp}. Such conversions taking place inside an external magnetic field could be extremely bright, even if a small percentage of the gravitational-wave pulse is transformed \cite{Gertsenshtein:1962,Fargion1995}. Modern calculations include considerations about plasma frequency and QED corrections \cite{Dolgov2017}, see also \cite{PhysRevD.99.044022,Raffelt:1987im}. As mentioned in \cite{Dolgov2017}, these articles dealt with the graviton-photon conversion at high frequencies $\omega/(2\pi)$ which exceeded the plasma frequency, $\omega_p/(2\pi)$, of the surrounding and intervening medium. It makes sense that higher frequencies were considered since low-frequency waves do not propagate in plasma. However, for the case of LIGO events concerning the merger of two black holes of roughly 20 - 30 $M_{\odot}$, the realistic gravitational-wave frequencies were approximately 100 - 200 Hz. This is problematic since typically the plasma frequency is approximately equal to 10 kHz$\sqrt{\frac{n_e}{1 \, \mathrm{cm}^{-3}}}$, where $n_e$ is the electron number density.
	
	In \cite{Dolgov2017}, they consider the possibility that the energy transition from gravitational-waves into electromagnetic waves may still be possible despite their low frequency. In particular, they show that gravitational-waves travelling through a high-frequency plasma and non-zero magnetic field continue to transform some of their energy into non-propagating plasma waves which heat up the surrounding plasma, thus leading to a noticeable release of electromagnetic radiation. In this work, they use an asymmetric conversion regime for which conversion from photons into gravitons is less probable given that the wave vector for this solution is purely imaginary, corresponding to the damping of the electromagnetic wave travelling in the plasma with frequency higher than that of itself. The authors conclude that the graviton-photon conversion mechanism studied can hardly account for plasma heating in LIGO black hole merger events.
	
	A further interesting paper is that of \cite{PhysRevD.99.044022} where they considered the effects generating dispersion and coherence braking of the electromagnetic waves. Therein they obtained the energy power and energy power fluxes for quasi-perpendicular external magnetic fields in the gravitational-wave propagation direction. The authors found that the energy power was large, but that the fluxes seen on Earth would remain faint. They considered waves with $\omega > \omega_p$, making the graviton-photon mixing for gravitational-waves of less than a few hundred Hertz less appealing. They noted, however, that the calculated plasma frequency through which the waves travel depends on the line of sight, producing varying frequency cut-off's. Finally, they conclude that the detection of this graviton to photon mechanism is unlikely to be made on Earth or in interplanetary space due to the large cut-off frequencies. They suggest that detection is only probable outside of the solar system. We show that this is not necessarily true.
	
	Our proposal uses the idea that gravitational-waves may be converted into electromagnetic waves following the mechanism laid out in \cite{Raffelt:1987im}. Extending this mechanism, beyond the treatment of axions, to gravitons implies that the gravitational field is quantized. Therefore, this model also provides a potential way to probe quantum gravity and its associated energy scales. Using the characteristics of relatively well studied environments, such as the known external magnetic field, plasma density, and a phenomenological energy scale $M$, we can compute the transition probability that gravitons are converted to photons by numerically solving the equations of motion laid out in \cite{PhysRevD.99.044022}. We then make use of data on the magnetic field and plasma environment within the nearby galaxy M31 to compute transition probabilities between photon and graviton when a graviton flux passes through the M31 galaxy. In particular, as the conversion spectrum from a single merger event will be shown to be likely unobservable, we will also study the observability of conversion from an isotropic background formed by neutron star mergers out to redshift 6 by using population synthesis estimates of the merger rate for binary neutron stars~\cite{Mapelli_2018}. 
	
	Supposing that it may be possible to find the parameter $M$ through observation of the electromagnetic spectrum corresponding to the gravitational-wave event, we may then compare its value to the phenomenological $M_{Pl} = \sqrt{\hbar c/G} = 1.22 \times 10^{19}$ GeV/c$^2$, the Planck mass, as is widely used in the literature as an expected energy scale for the appearance of quantum gravitational effects. In this way, a bound may be placed on any deviation of the energy scale of quantum gravitational effects $M$ from that of the Planck mass. Furthermore, such an observation could determine the existence of the graviton and therefore that the gravitational field is quantized. 
	
	Our means of analysis is to make a phenomenological extrapolation of a numerically determined gravitational-wave spectrum produced by a binary neutron star merger. We then determine how much the detectability of the resulting counterpart photon spectrum depends upon the parameters of the extrapolation, which is justified in that it accounts for both scenarios encountered in the literature: mode decay and the contribution of higher order modes extending the spectrum to higher frequencies. If the dependence is strong we conclude that detection is improbable without significant fine-tuning of the extrapolated spectrum. On the other hand, if the counterpart's detectability is largely independent of the choice of parameters, we take this to indicate that detection would be a strong possibility: in that, the extrapolation to detectable frequencies can be made far more naturally. To make the observations mentioned above, we consider a radio array on the far side of the moon and consider the cases: 100 and 1000 antennas. The far side of the moon is shielded from the interference from Earth and is thus preferable for conducting very low-frequency radio experiments. Considerations for building telescopes on the moon are already under way and have been considered in the literature \cite{BURNS2012433,KLEINWOLT2012167,bergman2009explorer,2013EPSC....8..279B,olfar,Rajan_2015,dsl,dsl1}. Furthermore, we consider the transition probabilities required for a detection with the low-frequency Square Kilometre Array (SKA-LOW). Although the first attempts to search for these counterpart signals can be performed with the SKA-LOW, we find that a detection of single merger events is probably impossible and, for an unresolved background of mergers, would require a considerably fine tuned behaviour with regards to the extrapolation of the gravitational-wave spectrum from neutron stars merger. This is largely due to the fact that the transition probabilities required for detection with the SKA-LOW are too high to be supported by any well known environments. Very large magnetic fields, in addition to large intervening structures would be required. However, reasonable transition probabilities result in a more achievable measurement using a lunar array. In this work, we demonstrate that, for an isotropic background formed by binary neutron star mergers, graviton-photon conversion emissions in the radio band are potentially detectable for the transition probabilities induced within the M31 environment provided exponential mode damping dominates only for frequencies above 80 kHz, an order of magnitude above the highest frequencies calculated in \cite{Kawamura:2003hu,Kawamura:2004ah,Tsang_2019}. This suggests that detection of an unresolved background may be highly challenging but is not impossible. We note that \cite{Gertsenshtein:1962} discuss the idea of the observability of a single merging event as well, but this is confirmed here to be improbable. If the converted EM spectrum is not detected by the hypothetical lunar array, and simulations do not indicate a cut-off before 80 kHz, we may also place a lower limit on the size of the scale $M$ by considering the largest $M$ attainable when detection does not require significant fine-tuning. We also compare our results to Lorentz invariance violation constraints and find that we produce competitive limits with respect to the results in \cite{Vasileiou:2013vra}. These limits, however, are contingent on the spectral extrapolation meeting the requirement of a minimum cut-off frequency value at $80$ kHz and a slope $> -2$. The 80 kHz cut-off receives some potential justification from the fact that higher-order harmonics in binary mergers can result in power-law spectral tails and even, in some circumstances, MHz frequency waves~\cite{Berti:2009kk,Martinez:2020cdh}.
	
	A vital issue will be that of background radio emissions~\cite{Dowell:2018mdb} and of the foreground emissions from the conversion environment itself. This issue, as discussed in \cite{2009NewAR}, in the context of reionisation science, will mean that it is vital to characterise the background/foreground emissions in detail in order to extract the signal due to gravitational-wave conversion. One approach that we discuss is that of using a distant galaxy as a conversion environment, which allows the expected signals to be only lightly impacted while minimising the issue of the foreground. 
	
	This paper is structured in the following manner: After the introduction,  section \ref{sec:level1} discusses the model we use to formulate our study. Section \ref{moon telescope} considers the construction of a lunar array in an optimal setting. Section \ref{r&d} presents our results and discussion with the conclusion presented in section \ref{con}.

	\section{Model}
	\label{sec:level1}
	
	Our model uses the spectrum for gravitational-waves resulting from a merging binary neutron star system found in \cite{Kawamura:2003hu,Kawamura:2004ah}.  In this model the stars have equal masses of 1.5 $M_{\odot}$ at a separation of 42.6 km. The resulting remnant is a low-mass black hole of about  2.8 $M_{\odot}$. The total energy radiated away as gravitational-waves, in less than 3 ms, is $\Delta E_{\mathrm{gw}} \sim 3 \times 10^{51}\; \textnormal{erg}$.
	
	Now let us suppose that part of the radiation emitted as gravitational-waves, is converted into electromagnetic radiation via the conversion of gravitons into photons, as described below in Section \ref{sec:level2}. We denote this fraction by $p= E_{\mathrm{EM}}/E_{\mathrm{total}}$, called the conversion probability, or conversion fraction. Then $E_{\mathrm{total}}=E_{\mathrm{gw}} + E_{\mathrm{EM}}= (1-p)^{-1}E_{\mathrm{gw}}$ where $E_{\mathrm{gw}}$ is the energy that remains in the form of gravitational waves after conversion and $E_{\mathrm{total}}$ is the initial gravitational wave energy of the event.
	
	
	\subsection{The mechanism}
	\label{sec:level2}
	
	Here we summarize some of the details of the conversion mechanism described in \cite{Raffelt:1987im}, which illustrates the possibility for a low-mass (or zero-mass) particle to be created from a photon (spin-1) passing through an external magnetic field, and vice versa. This formalism is applicable to the case of gravitons, which have spin-2. Conversion requires the presence of an external magnetic field supplying one virtual photon in order to satisfy symmetry constraints, thus conversion of real gravitons to real photons is one-to-one, as the second photon in this interaction is virtual.
	
	In \cite{PhysRevD.99.044022} the authors demonstrate an equation of motion for propagating photon and graviton mixtures, when we choose our coordinate system such that the propagation direction aligns with the $z$ axis, of the form
	\begin{equation}
	\left(\omega + \partial_z\right) \psi + \mathcal{M}\psi = 0 \; , \label{eq:eom}
	\end{equation}
	where $\partial_z = \frac{\partial}{\partial z}$, $\psi = \left(h_\mathrm{x}, h_+, A_x,A_y\right)^T$ with $h_\mathrm{x}$, $h_+$, $A_x$, and $A_y$ being graviton and photon polarisation states respectively. The mixing matrix $\mathcal{M}$ is defined as
	\begin{equation}
	\mathcal{M} = \left(\begin{array}{cccc} 0 & 0 & -i M_{g\gamma}^x & i M_{g\gamma}^y \\
	0 & 0 & i M_{g\gamma}^y & i M_{g\gamma}^x \\
	i M_{g\gamma}^x & -i M_{g\gamma}^y & M_x & M_{\mathrm{CF}} \\
	-i M_{g\gamma}^y & -i M_{g\gamma}^x & M^*_{\mathrm{CF}} & M_y \end{array}\right) \; , \label{eq:mix}
	\end{equation}
	where $M_{g\gamma}^q = g_{g\gamma} k B_q/(\omega + k)$ represents the photon-graviton coupling in the direction $q$ with $g_{g\gamma}$ being the coupling strength, $B_q$ being the magnetic field component in the $q$ direction, $\omega$ being the total particle energy, and $k$ being the momentum. $M_{\mathrm{CF}}$ contains the Faraday and Cotton-Mouton effects, and $M_q$ terms represent the `effective mass' picked up by the photon while propagating in a plasma. Note that we follow \cite{PhysRevD.99.044022} in the definition of all the matrix elements; so the reader is referred there for further details. It is possible to express the coupling in terms of some energy scale $M$ such that $g_{g\gamma} = \frac{1}{M}$. Unless it is noted that we are treating $M$ as a free parameter we will assume $ M = M_{pl}$. Note that the conversion of graviton to photon is one-to-one, as such they have equal energies and frequencies.
	
	We solve Eq.~(\ref{eq:eom}) numerically to obtain the state $\psi$ after propagating a distance $d$ in the environment of the M31 galaxy. The state $\psi$ will then be used to construct a probability $p (\nu)$ that a particle that starts as a graviton (with energy corresponding to $h \nu$) remains one after a distance $d$. Our initial state is taken to be composed entirely of gravitons so that $\psi_0 = \left(1/\sqrt{2},1/\sqrt{2},0,0\right)$ with the graviton terms chosen for simplicity. The solution method used will be to divide the magnetic field up into domains with a length sampled from a Kolmogorov turbulence distribution for coherence lengths between 50 and 100 pc~\cite{beckm31}. Each domain is assigned a value $B$ and plasma density $n_e$ according to the models below (using the coordinates of the centre of the domain) and is randomly allocated a magnetic field orientation. Each such allocation of field orientations to each domain is termed a realisation, with the observable solution taken to be the average over the ensemble of realisations. This approach is necessary as the actual configuration of the magnetic field is unknown, and the relative orientation influences the mixing matrix in Eq.~(\ref{eq:mix}). The averaging over the turbulent realisations will also remove any resonant spectral features.
	
	The M31 environment is characterised by magnetic field and the gas distributions for photon-graviton mixing, we will obtain these by following the model of \cite{ruiz-granados2010}. This means that our magnetic field strength is $B = 4.6 \pm 1.2$ $\mu$G at $r = 14$ kpc from the centre of the galaxy, with a more general profile following
	\begin{equation}
	B(r) = \frac{4.6 r_1 + 64}{r_1 + r} \, \mu\mathrm{G} \; ,
	\end{equation}
	where $r$ is the distance from the centre of M31 and $r_1 = 200$ kpc is taken to follow the more conservative value from fitting in \cite{ruiz-granados2010}. This was found to be valid when $r \leq 40$ kpc in \cite{ruiz-granados2010}. When extrapolating the magnetic field beyond $40$ kpc we will use the following conservative choice of profile
	\begin{equation}
	B(r) = \frac{4.6 r_1 + 64}{r_1 + 40\,\mathrm{kpc}} \exp(-(r-40\,\mathrm{kpc})/(3.8 r_d)) \, \mu\mathrm{G} \; ,
	\end{equation}
	where the scale radius $3.8 r_d$ is chosen motivated by arguments around the scale of spiral galaxy magnetic fields in \cite{beck2015}. 
	We take the gas density to be given by an exponential profile
	\begin{equation}
	n_e (r) = n_0 \exp\left(-\frac{r}{r_d}\right) \; ,
	\end{equation}
	where $n_0 = 0.06$ cm$^{-3}$ is the central density~\cite{beckm31}, and $r_d \approx 5$ kpc is the disk scale radius fitted by \cite{ruiz-granados2010}.
	
\begin{figure}[ht!]
		\centering
		\resizebox{0.7\hsize}{!}{\includegraphics{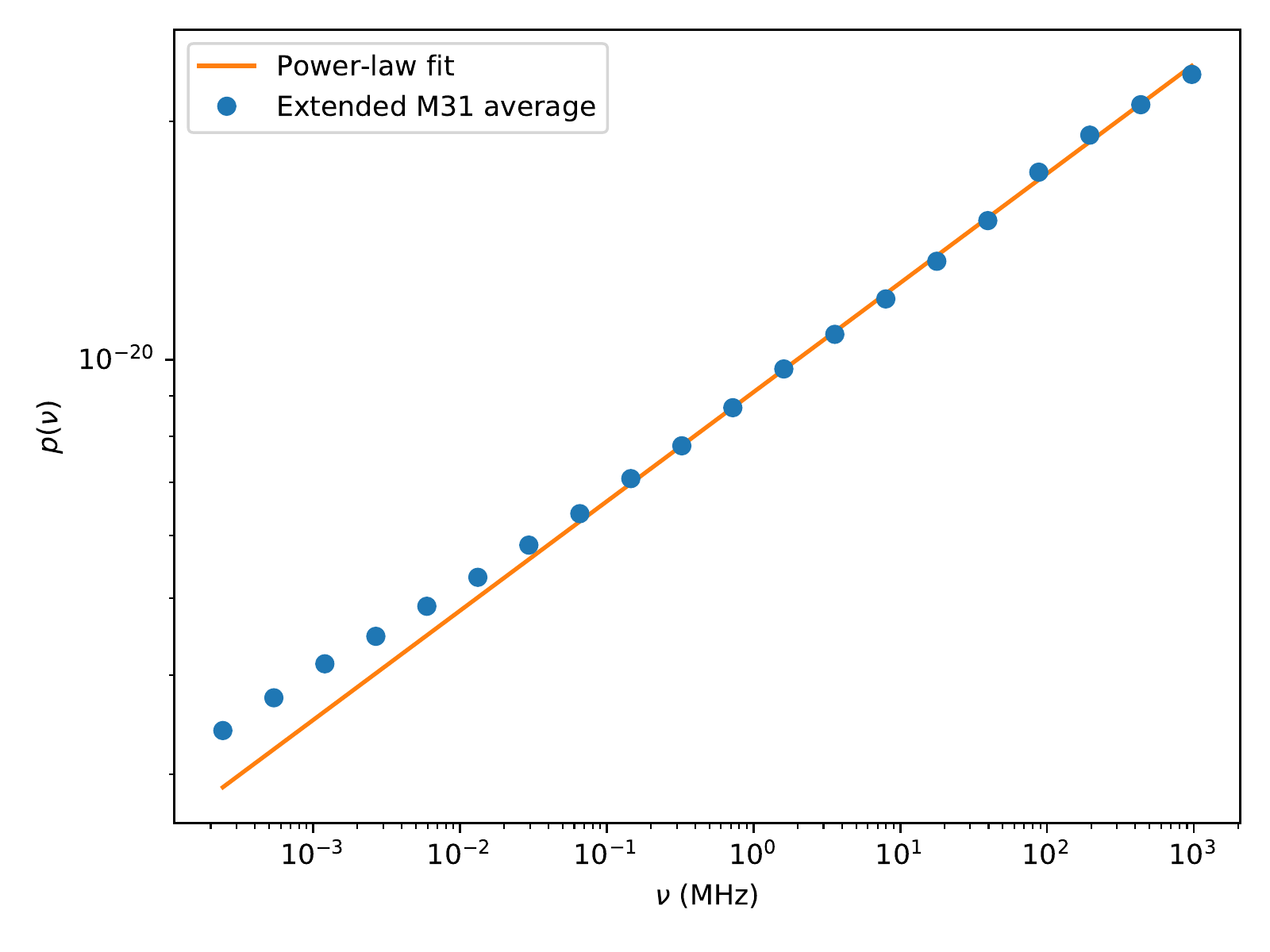}}
		\caption{Power-law fit to the conversion probability $p$ in the extended scenario from Table~\ref{tab:power-laws}.}
		\label{fig:powerlaw}
	\end{figure}

	We fit approximate power-law functions to the results of solving Eq.~(\ref{eq:eom}), one is fitted to $p(\nu)$ in the core region of M31 (when $r \leq 40$ kpc as studied by \cite{ruiz-granados2010}), the second is fitted to an integration over an extended region of M31 out to $r = 200$ kpc. The power-law takes the form $p(\nu) = a \left(\frac{\nu}{1\,\mathrm{MHz}}\right)^b$ and the parameters are listed in Table~\ref{tab:power-laws}. An example of the power-law fit for the extended case is displayed in Fig.~\ref{fig:powerlaw}, the deviation at low frequencies is a result of more complex shapes for $p(\nu)$ in the outer regions of M31 where the electron density drops faster than magnetic field strength, which can enhance conversion rates. In simple scenarios without multiple magnetic field domains or turbulence it is expected that, in the regime of weak mixing between graviton and photon and at low frequencies, the conversion probability scales as a steep power-law with index $=-2$ in a manner similar to the axion case~\cite{Horns:2012kw}. However, this mitigated by the radial variance of magnetic fields and plasma densities considered here, as well as the turbulence and domain structured magnetic fields.
	\begin{table}[ht!]
		\centering
		\begin{tabular}{|c|c|c|}
			\hline
			M31 region & $a$ & $b$\\ 
			\hline
			Core & $1.2 \times 10^{-20}$ & 0.15 \\
			Extended & $9.3 \times 10^{-21}$ & 0.13 \\
			\hline
		\end{tabular}
	\caption{Fitting functions for conversion probability $p$ between 1 kHz and 1 GHz.}
	\label{tab:power-laws}
	\end{table} 

	\subsection{The gravitational wave spectrum from a neutron star merger}
	
	The neutron star merger gravitational wave spectrum found in \cite{Kawamura:2003hu,Kawamura:2004ah} is calculated by simulating the contribution of up to 7 modes and reaches a frequency $\nu_f \approx 7$ kHz. We label this spectrum as $S_{\mathrm{gw}}(\nu)$. To reach potentially detectable frequencies we perform a power-law with exponential cut-off extrapolation for frequencies $\nu > \nu_f$ 
	\begin{equation}
	S_{\mathrm{ex}}(\nu) = A\nu^{\alpha}\exp\left(-\frac{\nu}{\nu_c}\right) \; ,
	\end{equation}
	so that the complete extrapolated spectrum is given by
	\begin{equation}
	S_{\mathrm{gw,ex}}(\nu) = \begin{cases}S_{\mathrm{gw}}(\nu) & \nu \leq \nu_f\\S_{\mathrm{ex}}(\nu) & \nu > \nu_f\end{cases} \; ,
	\end{equation}	
	\noindent where $\alpha$ is the power law index and $\nu_c$ is the cut-off frequency. The choice of using a power law with a cut-off may be explained as follows: a cut-off is applied because we know that the quasi-normal modes decay exponentially~\cite{Berti:2009kk}. Furthermore, we require a power law to account for the fact that higher order modes may extend the spectrum seen in \cite{Kawamura:2003hu,Kawamura:2004ah} as is done in \cite{Marklund:1999sp} in the presence of plasma and studied for MHz frequencies in \cite{Martinez:2020cdh}. Therefore, this extrapolation provides a parameter space of $\alpha$ and $\nu_c$ that effectively considers a large variety of scenarios for the extension of the gravitational wave spectrum to higher frequencies. The multi-order of magnitude extrapolation is justified in that it deals effectively with the cases of complete gravitational-wave mode decay (i.e. where $\nu_c \simeq \nu_f$) as well as the potential contribution of additional modes where $\nu_c > \nu_f$. It is well known that higher order modes contribute gravitational-wave power at higher frequencies but with shorter lifetimes~\cite{Berti:2009kk}. How much power is contributed by these modes is determined by the magnitude of their excitation, which is not yet well-understood for complex environments like neutron stars where many effects may amplify higher order mode magnitudes~\cite{Berti:2009kk,Martinez:2020cdh}. However, it is possible for even 1 MHz waves to be generated in eccentric systems as explored in \cite{Martinez:2020cdh}, this at least indicates that high-frequency waves are possible when higher harmonics are excited. Note that we do not require the same level of excitation as explored in \cite{Martinez:2020cdh} as we are not considering direct gravitational-wave detection at high frequencies, all we require is the potential for relatively weak excitation of higher order modes (as embodied by the power-law extrapolation). Additional justification for the power-law extrapolation can found in the spectral modelling performed in \cite{Tsang_2019}, where the authors determine the gravitational wave strain spectrum, from binary neutron star merger, out to 8 kHz. The aforementioned spectrum (figure 8 of \cite{Tsang_2019}) shows a power-law type decline at the highest displayed frequencies and no evidence of exponential decay below 8 kHz. Although a strain spectrum is not an identical quantity to the energy spectrum of \cite{Kawamura:2003hu,Kawamura:2004ah} it is sufficient to establish a lack of exponential decay in the gravitational-wave amplitude at frequencies studied so far in the literature. In this regard our extrapolation is reasonable as it assumes the smallest possible $\nu_c$ is 8 kHz, individual modes decay exponentially~\cite{Berti:2009kk} and it may be possible that higher order mode contributions cut off via an exponential decay. We use a power-law before this cut-off point as we note an overall trend of power-law like decrease in the amplitude of the gravitational-wave spectrum with frequency in \cite{Kawamura:2004ah} and the tendency of gravitational wave amplitudes from mergers to have power-law tails after the ring-down~\cite{Berti:2009kk}. By fitting to peaks of spectrum from \cite{Kawamura:2004ah} we determine that the amplitude of the spectrum declines as a broken power-law with frequency. At frequencies below $5$ kHz the slope is between $-3.5$ and $-3$ while it is $\gtrsim -1.8$ above $5$ kHz. This means that the vital range of extrapolation for the power-law spectral index can be taken as being between $-2$ and $-1.5$. 
	
	We normalise $S_{\mathrm{gw,ex}}(\nu)$ as follows: we require that $E_{\mathrm{gw}}$ is the gravitational wave energy emitted from the source at a luminosity distance $d_L$ over a time $\Delta t \approx 2$ ms and use this to determine the graviton flux at Earth.
		
	This power law is applied to the tail of the spectrum using the matching condition: $S_{\mathrm{gw}}(\nu_f) = S_{\mathrm{ex}}(\nu_f)$. Furthermore, $p(\nu) S(\nu)$ will then represent the spectrum obtained from the conversion of gravitational-waves into photons. The values $\alpha$ and $\nu_c$ provide us with the parameter space which allows us to determine the detectability of the electromagnetic counterparts with a radio telescope on the moon or with SKA-LOW. Note that this is a phenomenological extrapolation because it is difficult and beyond the scope of this work to calculate these spectra directly at frequencies high enough for radio-band detection. In  order that the exact details of the extrapolation don't bias the results, we will draw conclusions based on how insensitive the detection potential is to the choice of extrapolation parameters (the difficulties of even producing $S_{\mathrm{gw}}(\nu)$ are strongly noted in \cite{Kawamura:2003hu,Kawamura:2004ah}).
	
	\subsection{Isotropic gravitational wave background}
	
	We construct a gravitational wave background due to neutron star mergers as follows:
	\begin{equation}
	\Phi_{\mathrm{gw}}(\nu) = p(\nu) \Omega_{\mathrm{obs}} \int_{0.06}^{6} \mathrm{d}z \frac{\mathrm{d}V}{\mathrm{d}z} \tau(z) R_{\mathrm{merger}}(z) S_{\mathrm{gw,ex}}(\nu(1+z)) \; , \label{eq:phi-gw}
	\end{equation}
	where $\Omega_{\mathrm{obs}}$ is the solid angle being observed (in this case the sky-area of M31), $\frac{\mathrm{d}V}{\mathrm{d}z}$ is the co-moving volume element, $\tau(z)$ is the look-back time, and $R_{\mathrm{merger}}(z)$ is the rate of neutron star mergers per unit volume as found in \cite{Mapelli_2018}, the integral limits fit the plots presented in \cite{Mapelli_2018}. We note that black hole merger rates from \cite{Mapelli_2018} were found by the authors to be consistent with those of LIGO/VIRGO for the local universe, this has remained the case with further analysis and also demonstrates agreement for binary neutron stars~\cite{Abbott_2019,Abbott_2020}. It is vital to note that uncertainties in $\Phi_{\mathrm{gw}}(\nu)$ will be in direct proportion with those of $R_{\mathrm{merger}}(z=0)$ (as we could express $R(z) = R_0 r(z)$) and that these are substantial, as demonstrated by the fact that LIGO/VIRGO finds $R_0 \in [250,2810]$ Gpc$^{-3}$ yr$^{-1}$~\cite{Abbott_2020} while \cite{Mapelli_2018} find $R_0 \in [20,600]$ Gpc$^{-3}$ yr$^{-1}$. This implies, of course, that the flux from $\Phi_{\mathrm{gw}}(\nu)$ is uncertain by roughly a factor of 10. Notably the overlap between LIGO/VIRGO and \cite{Mapelli_2018} suggest the higher end of the range $R_0 \in [20,600]$ Gpc$^{-3}$ yr$^{-1}$ is more probable, which is promising for the detection prospects presented here. The cosmological quantities are calculated assuming the Planck results from \cite{planck2015}.
	
	We could, in principle, consider the isotropic background of gravitational waves passing through a succession of environments on the line of sight, thus leading to greater conversion fractions. However, to make a conservative estimate we consider only a single environment with some information as to its gas and magnetic field distributions. This is important as the evolution of these quantities with the host-environment redshift is largely uncertain.
	
	Note that, in our construction of $\Phi_{\mathrm{gw}}$, we have assumed that all the binary neutron star pairs and mergers are similar to that studied in \cite{Kawamura:2003hu,Kawamura:2004ah}. However, this assumption shouldn't be overly problematic as \cite{Mapelli_2018} indicate a strong concentration of binary neutron stars towards lower masses, like those studied in \cite{Kawamura:2003hu,Kawamura:2004ah}.
	
	In the M31 core region $\Omega_{\mathrm{obs}} = 4 \pi \times 6.73 \times 10^{-4}$ sr and in the extended case $\Omega_{\mathrm{obs}} = 4 \pi \times 1.55 \times 10^{-2}$ sr.

	\section{Lunar telescopes: Are they just science fiction?}
	\label{moon telescope}
	
	Presently, several large Earth-based radio telescopes have been built which operate at low-frequencies \cite{Chen_2018}. These include the Low Frequency Array (LOFAR) \cite{refId0} situated in Europe, the Long Wavelength Array (LWA) \cite{5109716} situated in New Mexico and the Murchison Wide-Field Array (MWA) \cite{Tingay_2013} situated in Western Australia. All of these only work above 10 MHz due to the opaqueness of Earth's Ionosphere \cite{Chen_2018}. Furthermore, below frequences of 30 MHz, radio observatories are severely limited due to Radio Frequency Interference (RFI) caused by GPS, cellphones, radio broadcasts, etc.
	
	This means that there is a whole range of frequencies below 30 MHz in the electromagnetic spectrum to which Earth-based experiments are blind. To explore this Ultra Long Wavelength (ULW) band, a space-based detector is required, placed above the Earth's Ionosphere. In a report by the European Space Agency \cite{report_1997}, they discuss the far side of the moon as a viable option for placing a very low frequency array. Published in 1997, this is the most comprehensive study on ULW radio observations to date~\cite{Chen_2018}. This idea is backed by the space-based instrument called the Radio Astronomy Explorer (RAE), launched in the 1970's. It discovered that the moon can act as a good shield against the RFI from Earth, thus providing an ideal radio-frequency observation environment.
	
	The development of space-based detectors is still largely in the conceptual phase. Here we list some of the many concepts that have been presented for exploring ULW.
	These concepts include the single satellite element projects such as the Dark Ages Explorer (DARE) \cite{BURNS2012433} and the Lunar Radio Astronomy Explorer (LRX) \cite{KLEINWOLT2012167}. Furthermore, there are many small satellites forming multi-element interferometers: Formation-flying sub-Ionospheric Radio astronomy science and Technology (FIRST) \cite{bergman2009explorer}, the Space-based Ultra-long wavelength Radio Observatory (SURO-LC) \cite{2013EPSC....8..279B} and the Orbiting Low Frequency Array (OLFAR) \cite{olfar,Rajan_2015}. Additionally, we mention Discovering the Sky at the Longest wavelengths (DSL) \cite{dsl}, which would conduct single antenna measurements as well as form part of a ULW radio interferometer \cite{dsl1}. Finally, studies using dipole antennas have been considered \cite{bergman2009explorer,2013EPSC....8..279B,BURNS2012433,olfar,KLEINWOLT2012167}.
	Whilst there are currently no telescopes on the moon, it is evident there is much activity in this field with major international institutions and agencies participating and collaborating to put a detector on the moon or in orbit. We consider here some optimal specifications for a lunar array, that may in principle, detect graviton-photon conversion counterparts to gravitational-waves.
	

	
	\subsection{Building a telescope on the moon}
	
\begin{figure}[h!]
		\centering
		\includegraphics[width=0.6\textwidth]{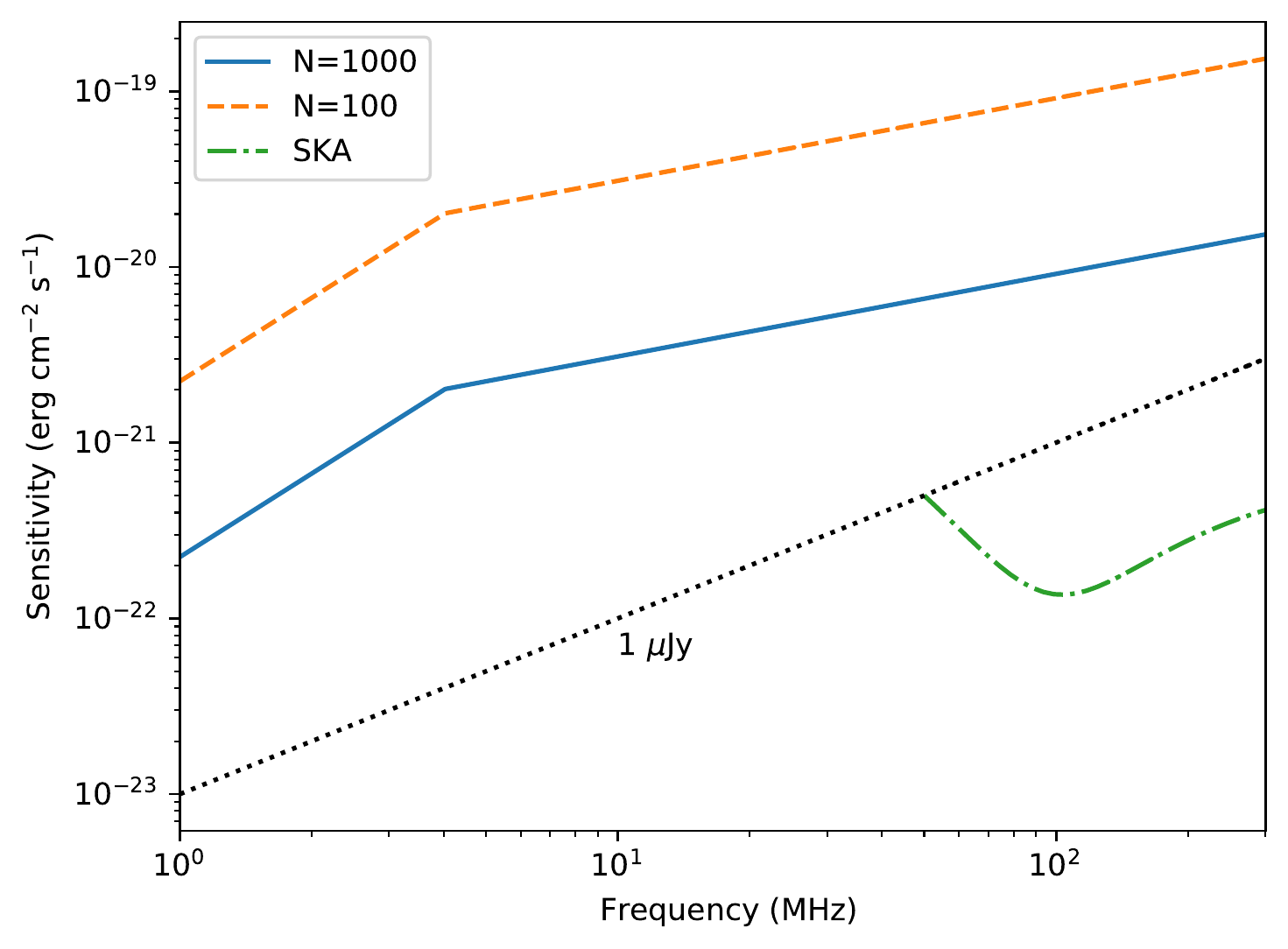}
		\caption{Minimum flux of photons that may be observed by a telescope on the moon within $100$ hours when there are $N=1\times 10^2$ antennas (orange/dashed), and for $N=1\times 10^3$ antennas (solid/blue) and finally we plot the sensitivity of the SKA-LOW (green/dashed-dotted) which has $N\sim 1 \times 10^5$ antennas. The dotted line displays a $1$ $\mu$Jy threshold. We extrapolate the lunar array sensitivity to frequencies above 30 MHz for comparison purposes.}\label{fig5}
	\end{figure}

	In this section we provide some details for designing an optimal lunar radio telescope. As a test case, we consider two set-ups: firstly, a configuration using $N = 10^{3}$ log-periodic dual-polarized dipole antennas with bandwidth $\Delta\nu=30$ MHz, and secondly, the case with $N = 100$ antennas and the same bandwidth $\Delta\nu$. The bandwidth is motivated by design considerations discussed in \cite{2009NewAR,Chen_2018}. We calculate the minimum observable flux for the array as follows\footnote{http://sci.esa.int/science-e/www/object/doc.cfm?fobjectid=53829}
	
	\begin{eqnarray}\label{eq3}
	S_{min} = \frac{2k_B T_{sky}}{N\sqrt{\Delta\nu\tau} A_e} \; ,
	\end{eqnarray}
	where $A_e = \lambda^2/4\pi$ is the effective collecting area and $\lambda$ is the incoming wavelength. Here $k_B$ is the Boltzmann constant, $\tau$ is the integration time. The sky temperature is given by \cite{2009newarray}
	\begin{eqnarray}\label{eq4}
	T_{sky} = 
	\begin{cases}
	16.3\times 10^6 \textrm{K} \left(\frac{\nu}{2 \mbox{MHz}}\right)^{-2.53},\;\; \nu > 2 \mbox{MHz}\\
	16.3\times 10^6 \textrm{K}\left(\frac{\nu}{2 \mbox{MHz}}\right)^{-0.3},\;\; \nu < 2 \mbox{MHz}\;\; .
	\end{cases}
	\end{eqnarray}
	We can see, in figure \ref{fig5}, that a lunar array would less sensitive than the SKA at 50 MHz achieving around 10 $\mu$Jy compared to $\sim 1$ for the SKA (note we have extrapolated the lunar array operating band for purposes of comparison with SKA). The curve, for a lunar array, has an expected break or kink in its uniformity due to the behaviour of equation (\ref{eq4}). In the aforementioned figure, we have computed both telescope sensitivities for a total integration time of 100 hours, to make comparison with standard figures quoted for the SKA~\cite{ska2012}. For a single merger event observation we would be limited to around a milli-second of integration time only, this being the event duration. For the lunar telescope we consider two cases, one for $N=100$ dipoles and the other for $N=1000$ dipoles. The SKA-LOW itself will possess $N \sim 10^5$ dipoles. The orange dashed line considers the conservative case of only $N=100$ dipole antennas, and as can seen we achieve better sensitivity in the case of $N=1000$ antennas given by the solid blue line. 
	
	\subsection{\label{section 3.2} Detectability of produced photons}
	
	In order for these counterparts to be detected by an observer on Earth or on the moon, the frequency of these photons must be larger than the plasma frequency, $\omega_p$, of the environments through which they travel. This is to ensure that the photons are not absorbed by the intervening medium. For a galaxy like M31 ($n_e = 0.06\times 10^{-3}$ cm$^{-3}$) we have $\omega_p \sim 1.54\times 10^{-2}$ MHz. For the same reasons, the photons must have frequencies higher than the plasma frequency of the atmosphere through which they travel to arrive at the detector. Whilst the moon has no atmosphere it does possess and Ionosphere with $n_e \sim 10^{2}$ cm$^{-3}$ \cite{1742-6596-653-1-012139} and therefore has $\omega_p\sim 0.62$ MHz and the Earth (n$_e \sim 10^{5}$ cm$^{-3}$) \cite{earthIonosphere}\footnote{http://solar-center.stanford.edu/SID/science/Ionosphere.pdf} has $\omega_p\sim 19.9$ MHz.
	
	The sensitivity profile for SKA-LOW may be found in \cite{ska2012}, whilst for the moon radio telescope we use the setup established in this section. The plasma frequency for the moon, as calculated above, is roughly $0.62$ MHz. Therefore, the telescope operating frequency should start just above this value, and we chose to start at 1 MHz. Hence, the operating frequency range is $1 - 30$ MHz.
	
	Note that the photons we are considering are those that \textit{do} strike the antennas after having passed through the moon's Ionosphere. Therefore, they have also survived travel through the intervening space between the moon and the gravitational-wave event.

	\section{\label{r&d}Results and discussion}
	
	We provide the parameter space which indicates values of $\alpha$ and $\nu_c$ for which detections may be possible within the bandwidths of SKA-LOW or a lunar array as described in section \ref{moon telescope}. That is, we study how the flux varies compared to the sensitivity of the detector. When the flux of the incoming photon is higher than the sensitivity of the detector, we have made a detection, the shaded regions of the parameter space are those in which detections are possible. SKA-LOW is represented by blue/dark shading and the lunar array by the green/light shading. Plots showing parameter space detection coverage were computed at a 5$\sigma$ confidence level and we will assume the energy scale of quantum gravitational effects is that of the Planck mass, making these results potentially conservative.
	
	With regards to our choice of the parameter space, we require $\nu_c$ to be larger than the final point in the spectrum in \cite{Kawamura:2003hu,Kawamura:2004ah}, but smaller than 100 MHz, above which $\nu_c$ becomes irrelevant for SKA-LOW. Ultra-steep radio-band power-law indexes have been shown \cite{DeBreuck:2000zk} to extend up until -2. In keeping with this, but allowing some lee-way for more extreme spectral fall-off, we chose our parameter space to lie within $\alpha\in [-3,0]$.
	
	As we make use of a phenomenological extrapolation of the gravitational-wave spectrum from \cite{Kawamura:2003hu,Kawamura:2004ah}, we will discuss the implications of the results qualitatively. We are interested in the breadth of the detectable parameter space, rather than the specific model values for which signals will be detectable. This is to ensure that our conclusions are not strongly dependent on the choice of extrapolation. So we will examine how strongly detection of a signal depends on the choice of model parameters. A weak dependence will be taken to indicate that the signal is highly detectable, as it can be naturally extrapolated to detectable frequencies with little or no fine-tuning. Whereas, a requirement of very particular values for $\alpha$ and $\nu_c$ will imply that detection is improbable, as significant fine tuning of the spectral extrapolation is needed to reach detectable frequencies. We make a reasonable apriori assumption that a detection requiring fine-tuning is less plausible the more fine-tuning that is required.
	
\begin{figure}[h!]
		\centering
		\includegraphics[width=0.49\textwidth]{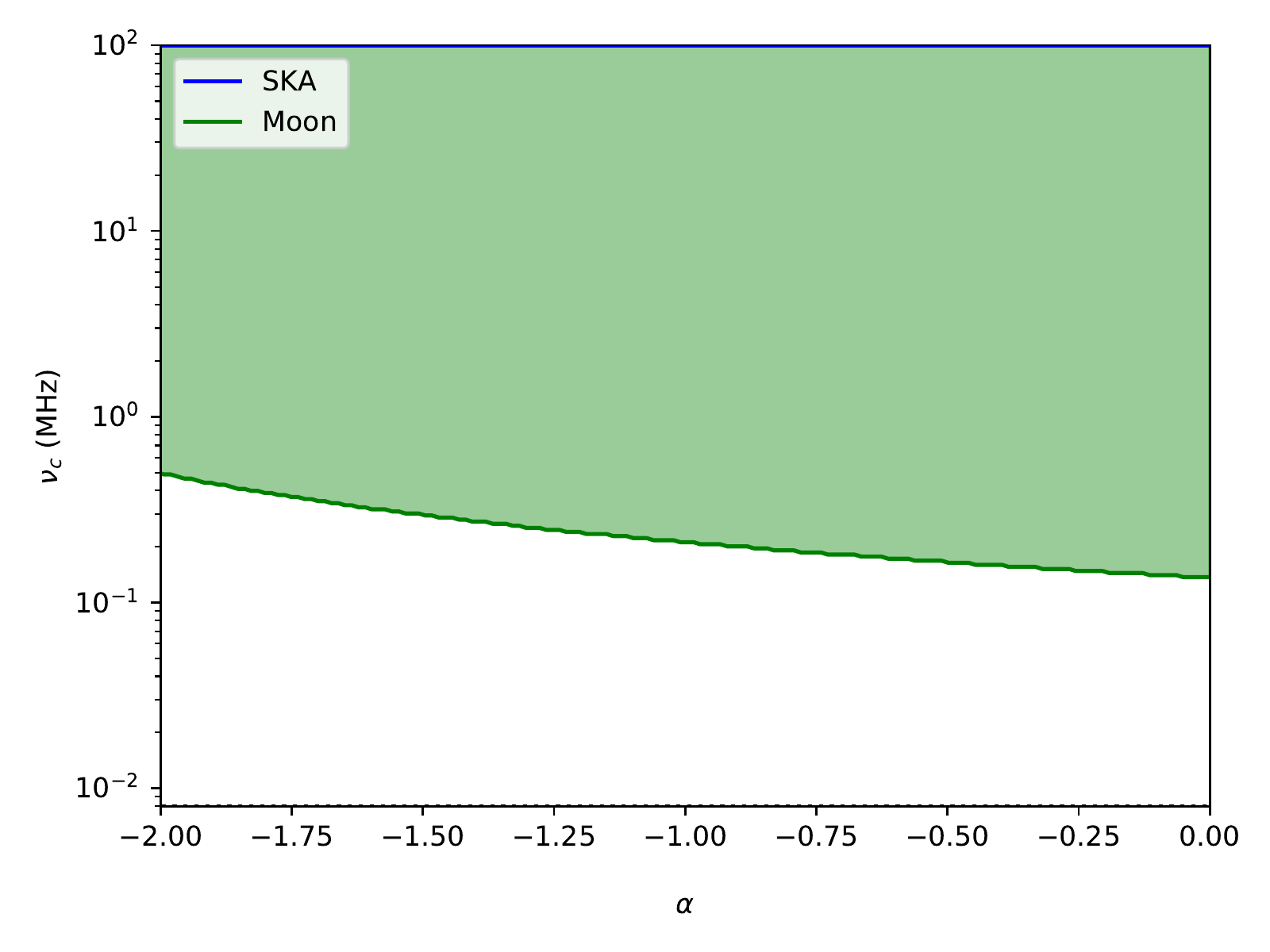}
		
		\includegraphics[width=0.49\textwidth]{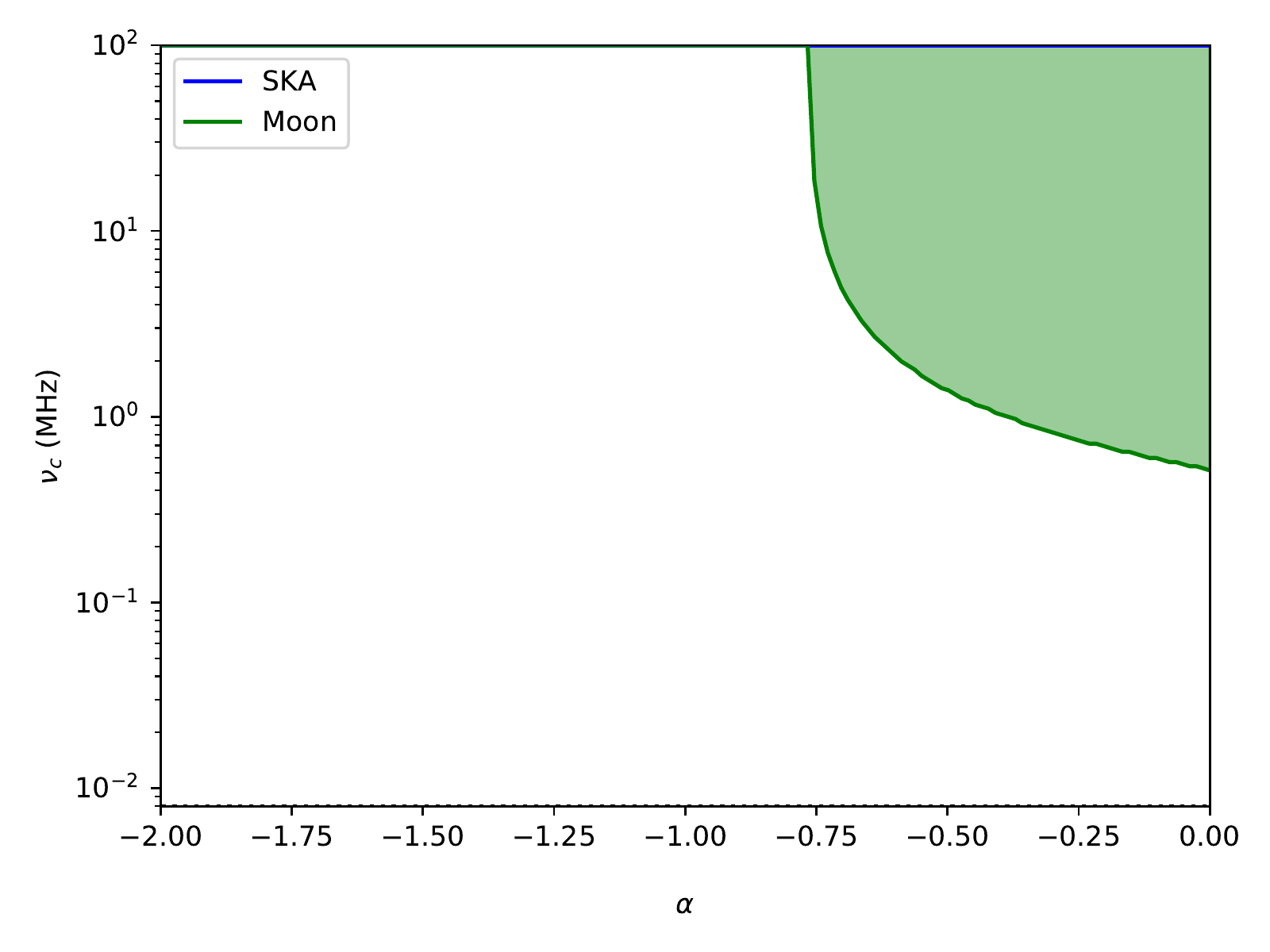}
		\includegraphics[width=0.49\textwidth]{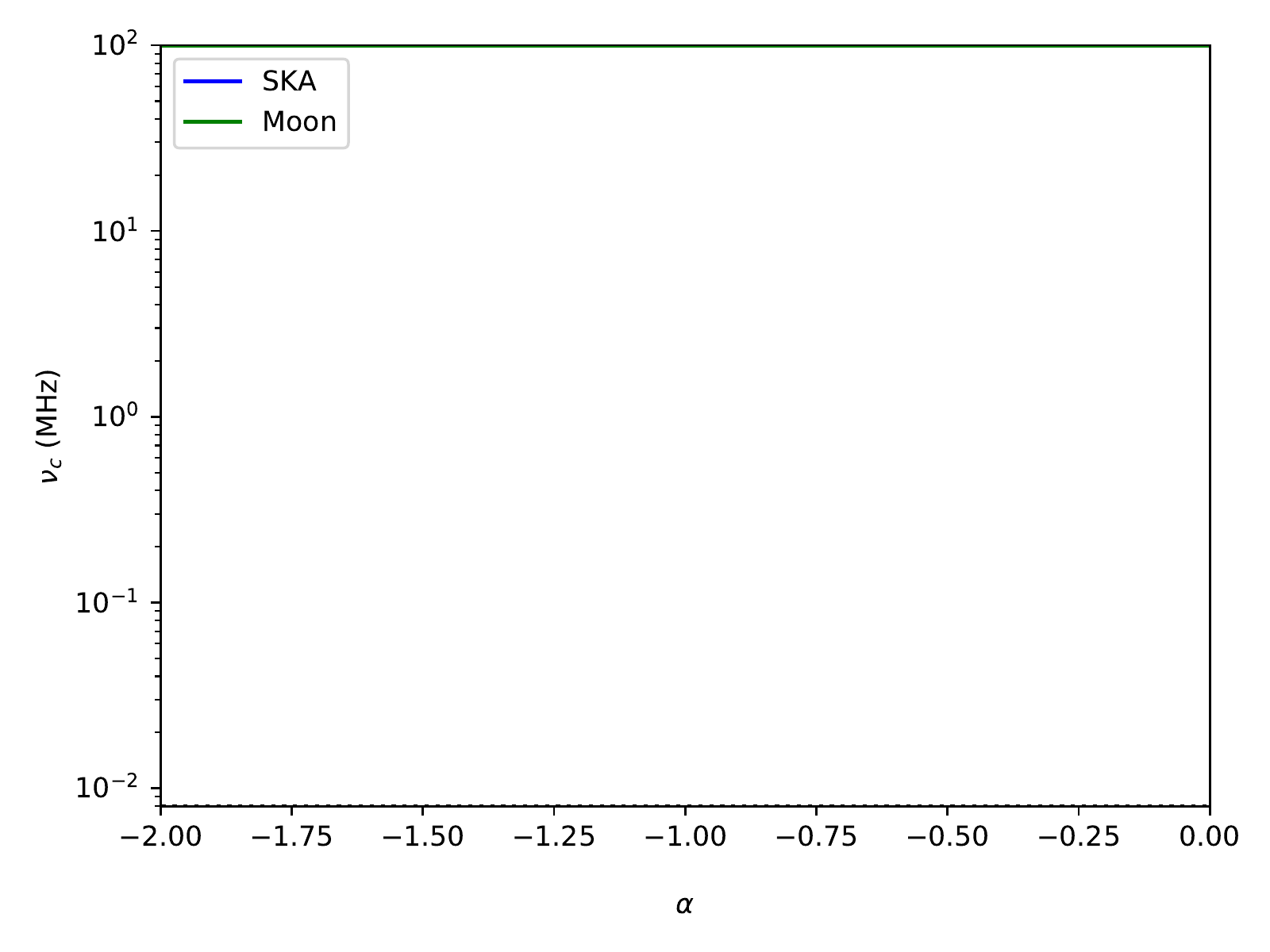}
		\caption{Detectability parameter space for the conversion of gravitons into photons in the M31 extended scenario with a single merger event. SKA-LOW is represented by the blue/dark region and the lunar array by the green/light region. Top: $10^5$ lunar antennae. Bottom left: 1000 lunar array antennae case. Bottom right: 100 antennae.}\label{fig:single}
	\end{figure}

	Figure~\ref{fig:single} displays the case of a single neutron star merger event occurring within the M31 galaxy. As is evident, the observation is only possible for more than 100 dipole elements in the lunar array and the parameter space coverage is weak with 1000 dipoles and struggles to reach the cut-off frequency of 80 kHz even with $10^5$ antennae. This suggests the detectability of the event is improbable, as it requires a very large array even if the gravitational wave spectrum only begins exponential decay more than an order of magnitude above the end of the spectrum from \cite{Kawamura:2003hu,Kawamura:2004ah,Tsang_2019}. In addition to this, the power-law extension cannot have $\alpha < -0.75$ for 1000 antennae. In an attempt to test a very close event we considered one taking place in the Milky-Way galactic centre assuming a magnetic field with an exponential profile and central strength of $60$ $\mu$G. This resulted in $p \approx 10^{-32}$, meaning that this cannot compensate for the flux boost of $\sim 10^4$ due to the smaller distance. We conclude that, barring some unconsidered nearby conversion environment, this makes it unlikely that single binary neutron star merger events can be observable via photon-graviton mixing without a very large array indeed. This result motivates our consideration of an isotropic background formed from a merging population of binary neutron stars instead.
	
\begin{figure}[h!]
		\centering
		{\includegraphics[width=0.49\textwidth]{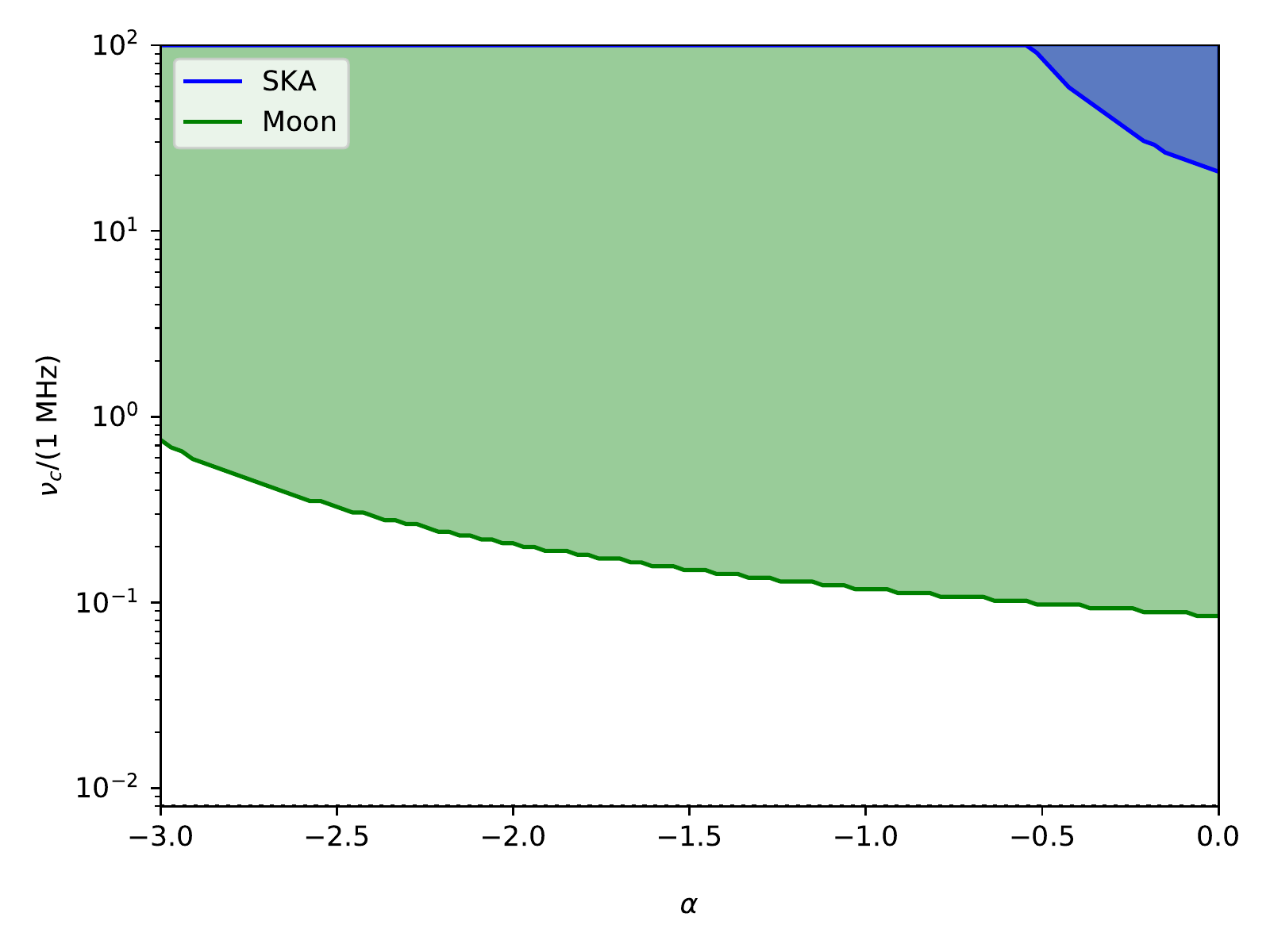}}
		{\includegraphics[width=0.49\textwidth]{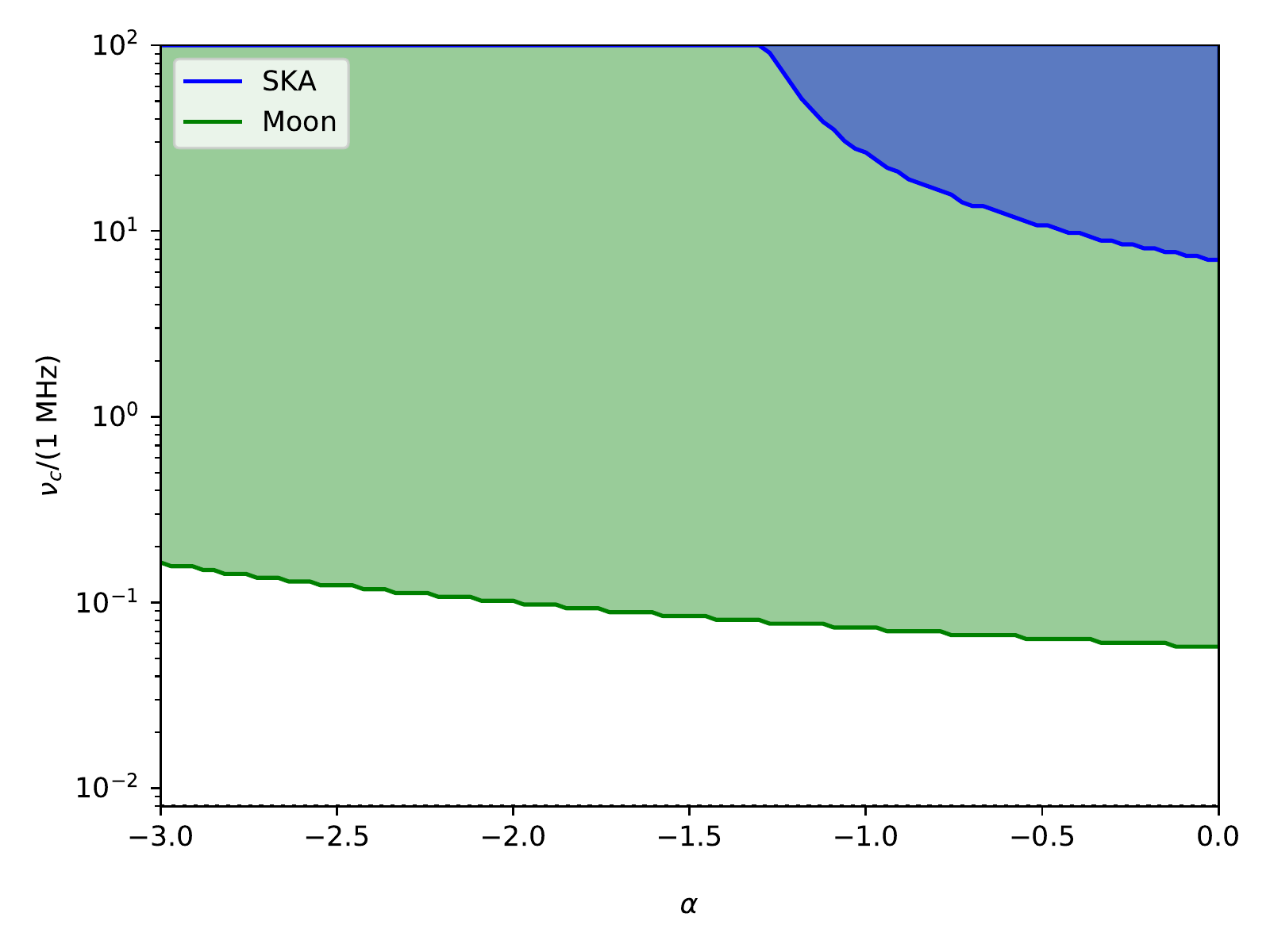}}	
		
		{\includegraphics[width=0.49\textwidth]{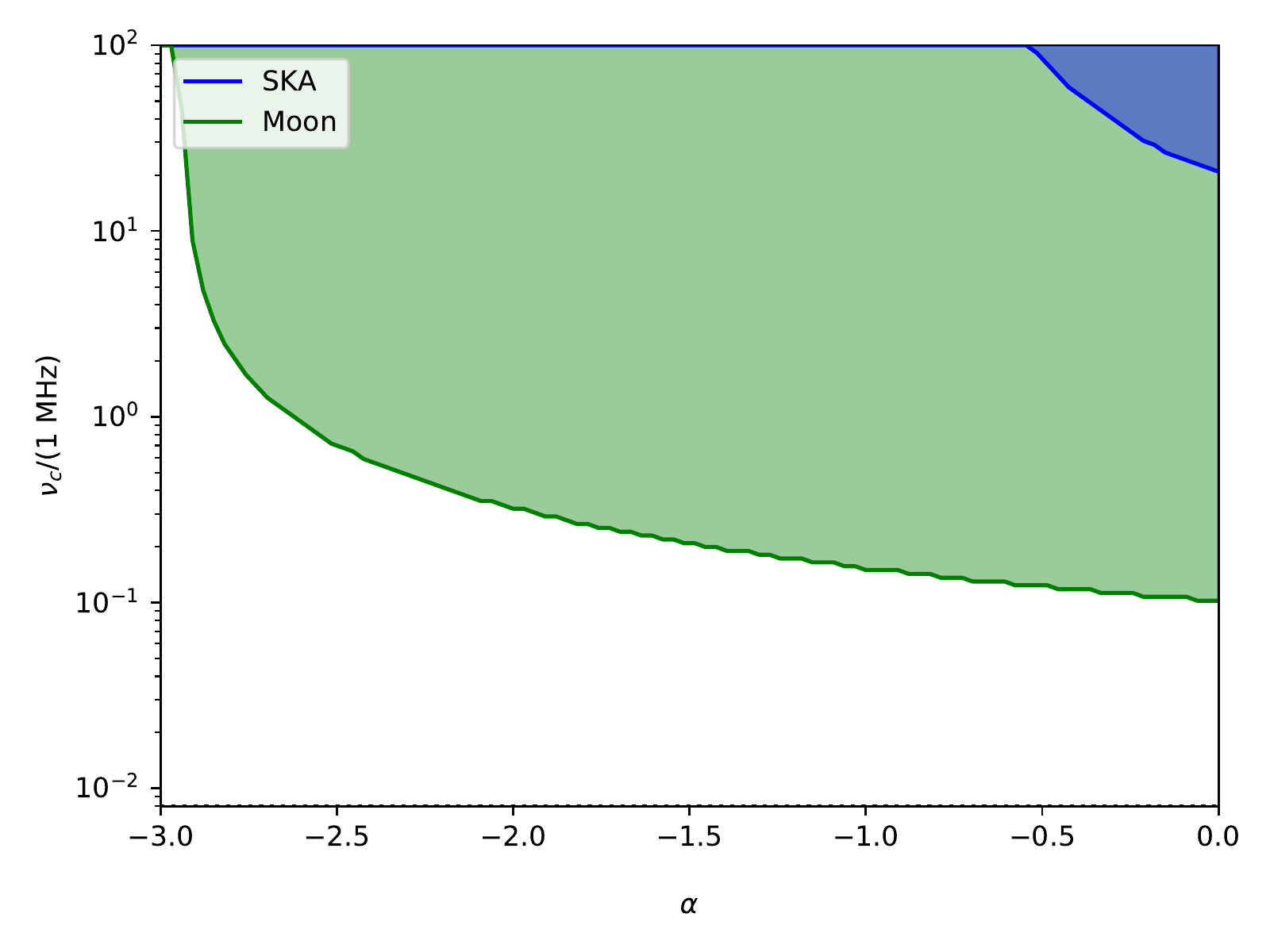}}
		{\includegraphics[width=0.49\textwidth]{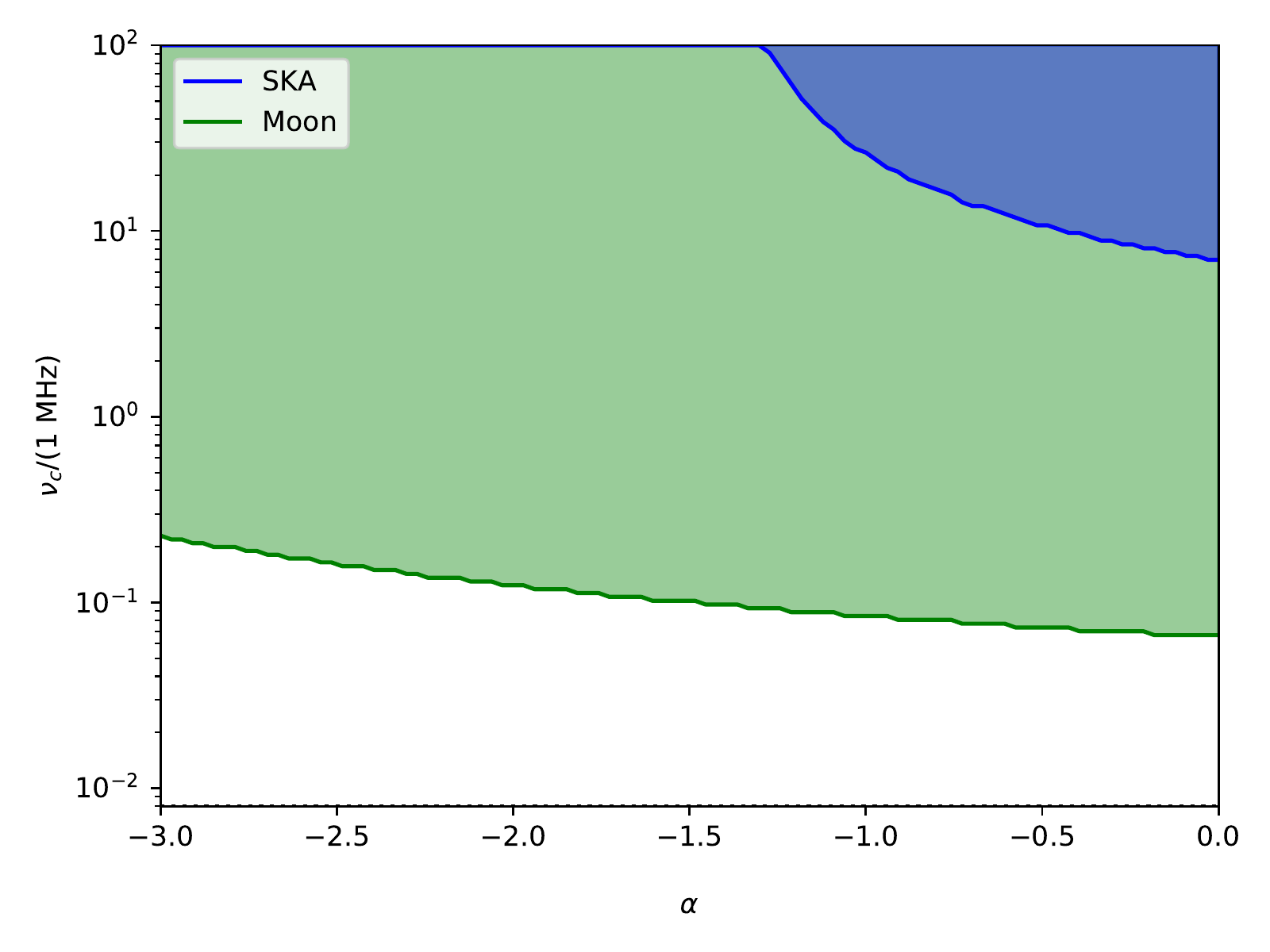}}	
		
		\caption{Detectability parameter space for the conversion of gravitons into photons in the M31 core scenario with an unresolved merger background. SKA-LOW is represented by the blue/dark region and the lunar array by the green/light region. Left: integration time of 1 s. Right: 100 hours of integration time. The upper row displays the 1000 lunar array antennae case and the lower row displays that with 100 antennae.}\label{figure2}
	\end{figure}

	Figure~\ref{figure2} displays the detectable parameter spaces using the M31 core propagation scenario for an isotropic gravitational wave background due to binary neutron star mergers. It demonstrates the effect of both the integration time, at 1 s (left column) and 100  (right column), as well as the number of lunar array antennae, 1000 (top row) and 100 (bottom row). These plots demonstrate the importance of the array size, as steeper power-law extrapolations are not so easily observable with the 100 element array and the extended integration time is important but only allows for approaching an order of magnitude above the highest frequency studied in \cite{Tsang_2019}. The coverage is not complete, even within an order of magnitude of \cite{Tsang_2019}, with cases where $\alpha < -2$ being unobservable if the cut-off starts at 80 kHz. The SKA, on the other hand, shows a somewhat limited detection potential, requiring the cut-off to be at least two orders of magnitude above the end of the spectra from \cite{Kawamura:2003hu,Kawamura:2004ah}. Although increased integration time is a considerable advantage over the single event results the real benefit comes from the combination with the flux increase from the large number of unresolved events. 
	
\begin{figure}[h!]
		\centering
		{\includegraphics[width=0.49\textwidth]{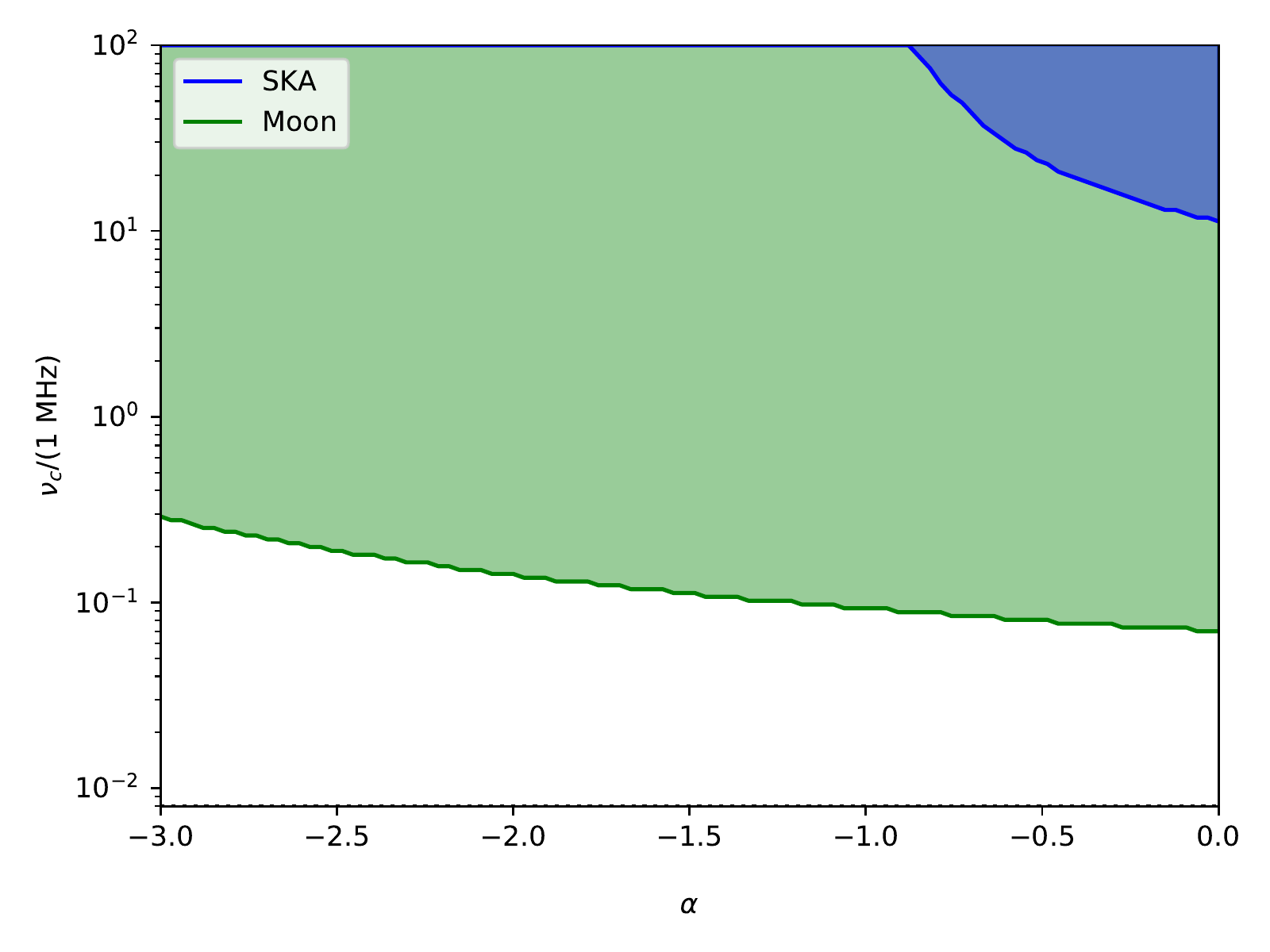}}
		{\includegraphics[width=0.49\textwidth]{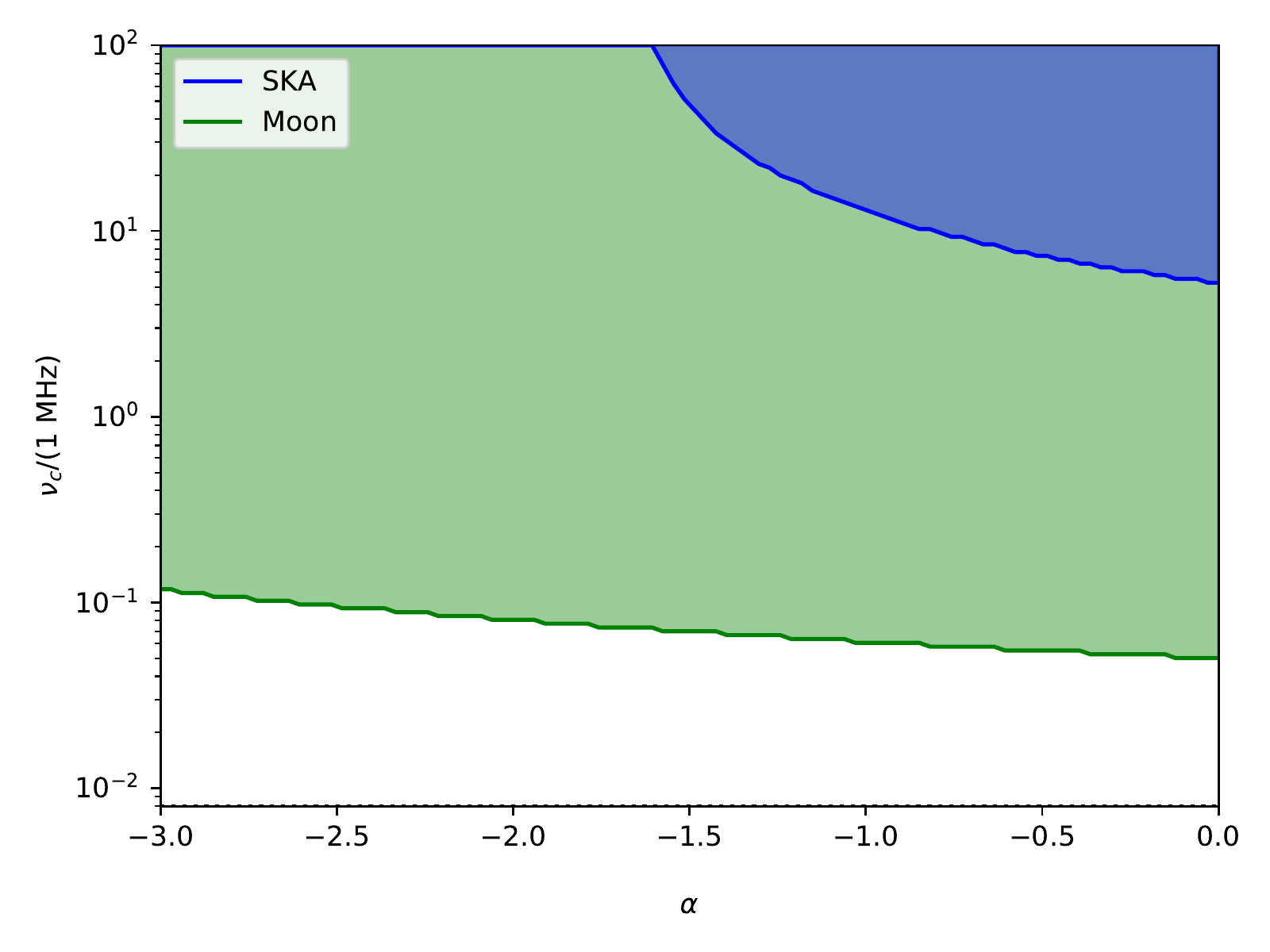}}	
		
		{\includegraphics[width=0.49\textwidth]{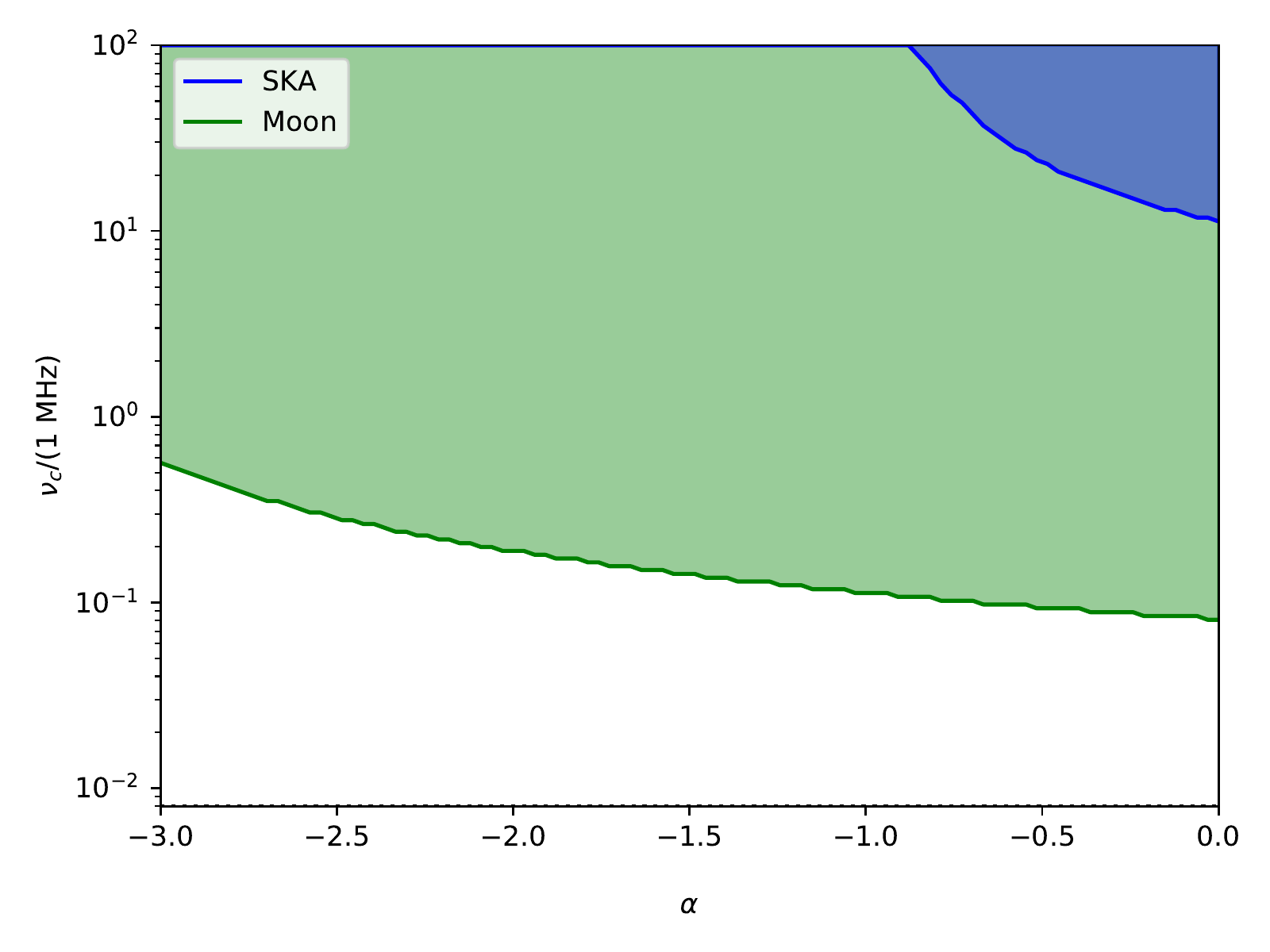}}
		{\includegraphics[width=0.49\textwidth]{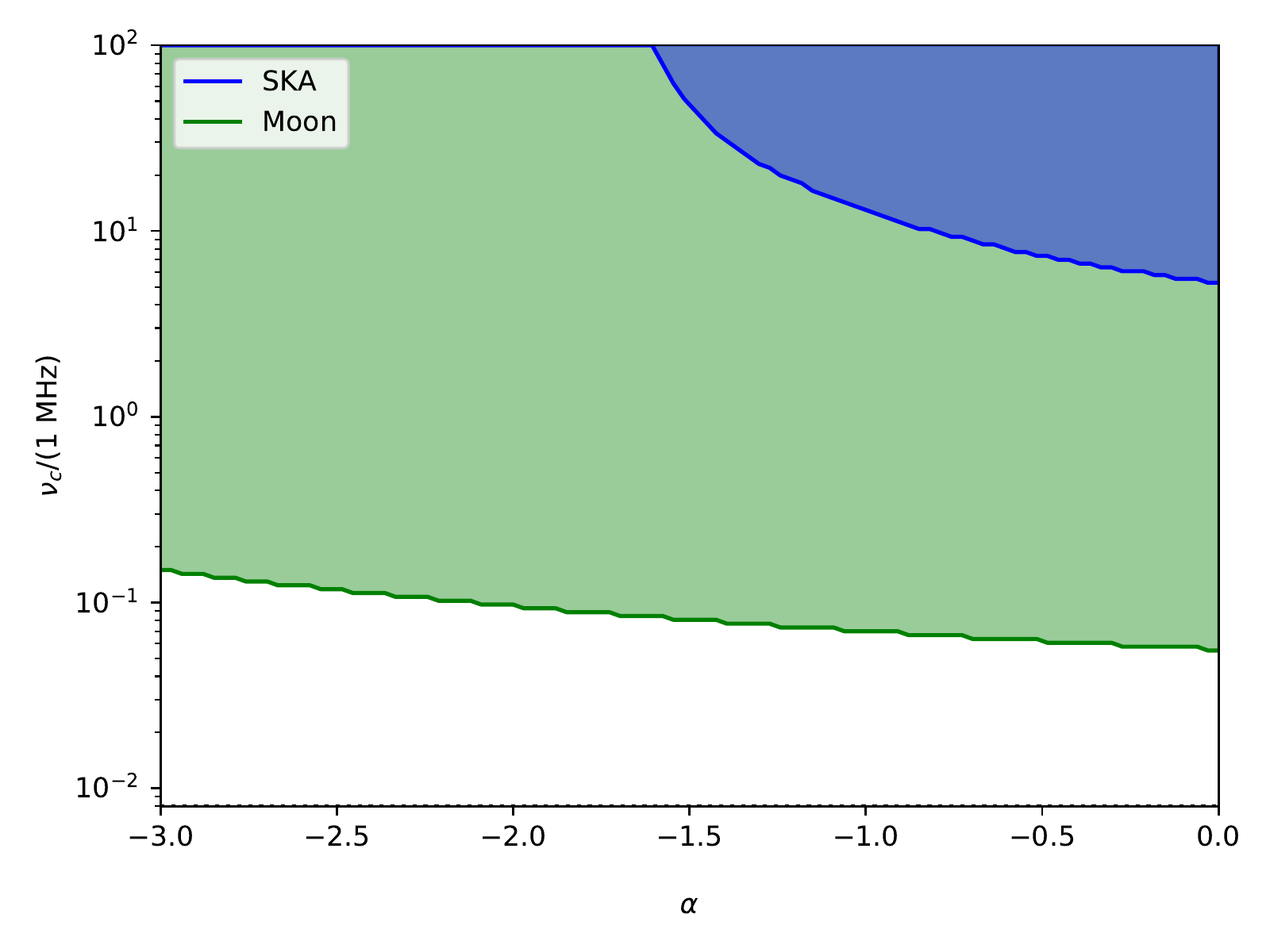}}
		
		\caption{Detectability parameter space for the conversion of gravitons into photons in the M31 extended scenario with an unresolved merger background. SKA-LOW is represented by the blue/dark region and the lunar array by the green/light region. Left: integration time of 1 s. Right: 100 hours of integration time. The upper row displays the 1000 lunar array antennae case and the lower row displays that with 100 antennae.}\label{figure3}
	\end{figure}

	Figure~\ref{figure3} displays a conservative choice for the extended M31 propagation scenario again for an isotropic gravitational wave background due to binary neutron star mergers. For cases with 100 hours integration and/or 1000 dipoles the results are very marginally superior to those from Fig.~\ref{figure2} as $p(\nu)$ is similar but the observed sky area is an order of magnitude larger. However, for smaller arrays and shorter integration times the detection prospects are improved substantially. The coverage of the parameter space is again incomplete, the best achievable cut-off frequency required for observation being an order of magnitude above the highest frequency explored in \cite{Tsang_2019}. This suggests that the controlling variable over the detectability is the cut-off frequency as it most strong dictates the parameter space coverage. Thus, the observability of the unresolved spectrum hinges how far the power-law trend at higher frequencies~\cite{Tsang_2019} in the merger spectrum continues. This means that detection is possible but there a large area of the extrapolation parameter space that could result in the signal being undetectable. We will supplement this by exploring how close to the Planck scale we can place stringent exclusion limits from the non-observation of such a GW-induced signal if the cut-off frequency can indeed be extended out to 80 kHz.
	
	\begin{figure}[!h]
		\centering
		\resizebox{0.8\hsize}{!}{\includegraphics{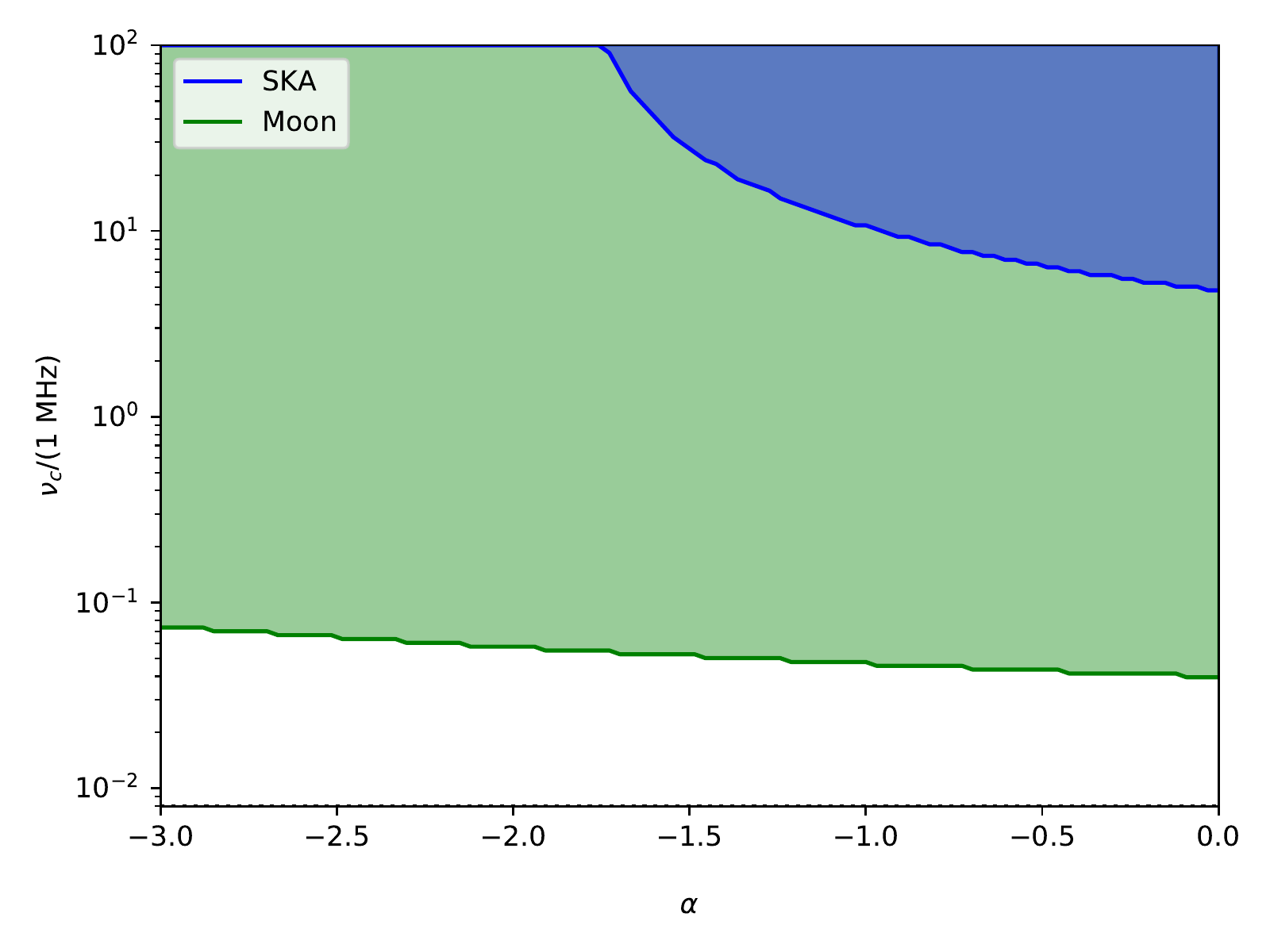}}
		\caption{Detectability parameter space for the conversion of gravitons into photons in the M31 extended scenario with an unresolved merger background. SKA-LOW is represented by the blue/dark region and the lunar array by the green/light region. The integration time is 1000 hours and the array contains 10$^5$ dipoles.}\label{fig:opt}
	\end{figure}

	Figure~\ref{fig:opt} conveys the most optimistic detection case: an array for $10^5$ elements (similar to that needed for 21 cm power spectrum observations~\cite{2009NewAR}) observing for 1000 hours. In this case we are no longer dependent on the power-law spectral index to reach $\nu_c \lesssim 80$ kHz.
	
\begin{figure}[h!]
		\centering
		\includegraphics[width=0.49\textwidth]{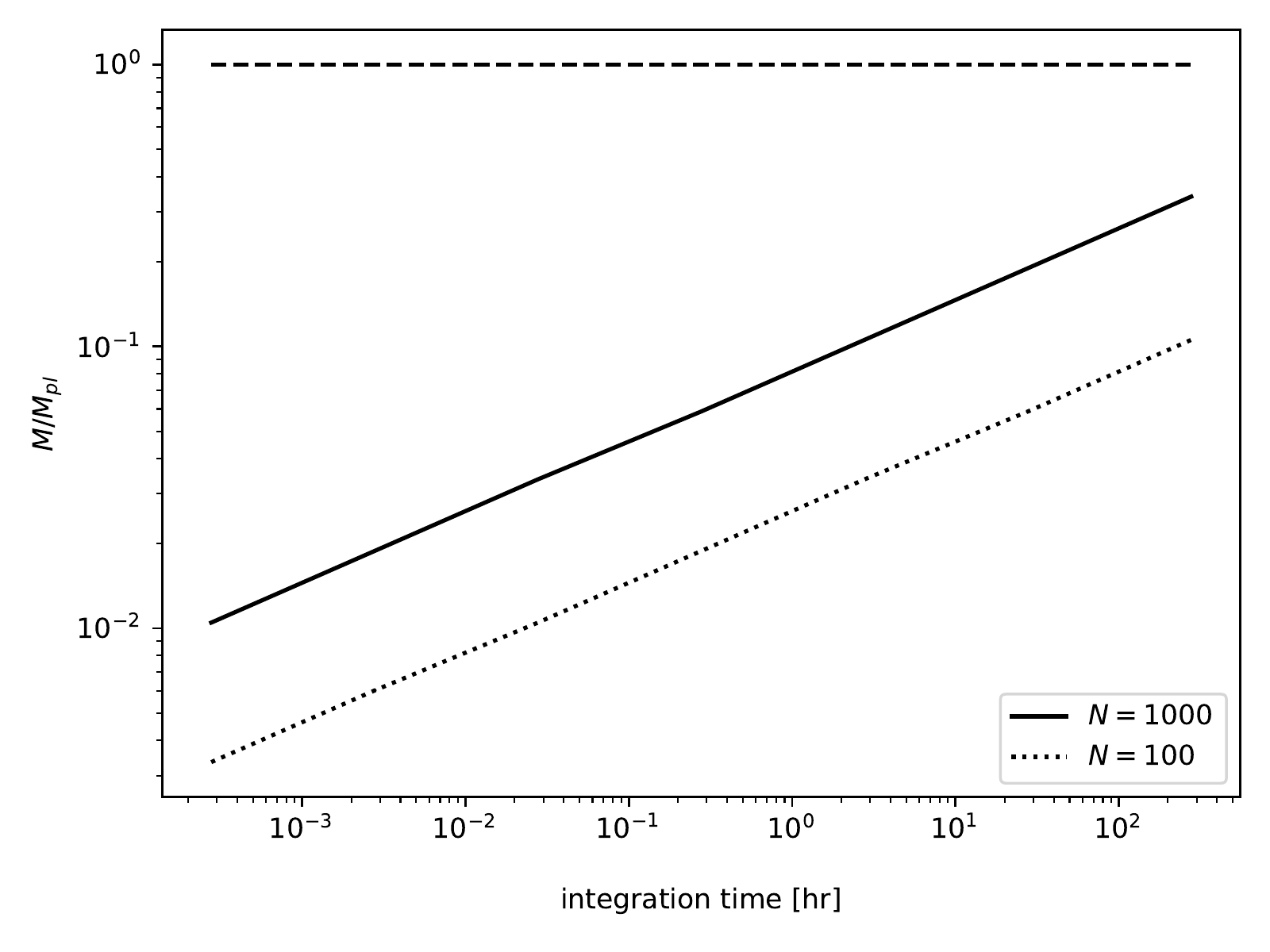}
		\includegraphics[width=0.49\textwidth]{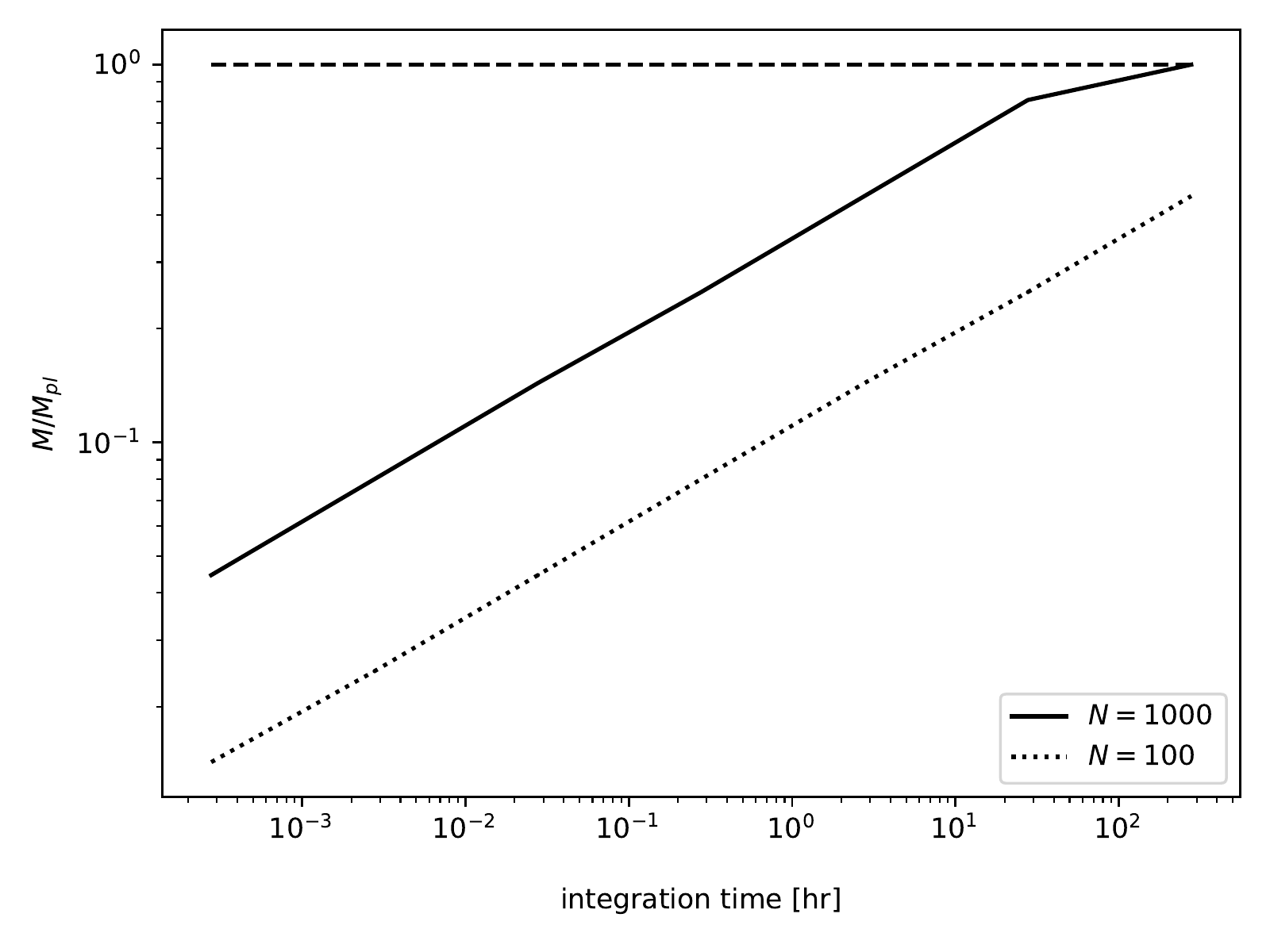}
		\caption{Maximum detectable energy-scale $M$, as a fraction of the Planck scale $M_{pl}$, versus telescope integration time when $\nu_c = 80$ kHz for an unresolved merger background. This plot illustrates the lower bounds on $M$ using a lunar array with $1000$ dipoles (solid lines) or 100 dipoles (dotted lines) and apply for $\alpha \geq -3$. Left: we assume the M31 core propagation scenario. Right: M31 extended case}\label{fig6}
	\end{figure}

	Figure \ref{fig6} shows the lower limit that can be placed on the scale $M$ via non-observation with the lunar radio array for an isotropic gravitational wave background due to binary neutron star mergers. We derive the limits by finding the largest $M$ observable at $2\sigma$ confidence interval in the most conservative, yet detectable, scenario where $\nu_c \approx 80$ kHz. The largest observable value will be a lower limit in the case of non-observation. An interesting comparison may be made with the Lorentz invariance violation limits for gamma-ray bursts. The Lorentz invariance violations can be parameterised by $E^n$, where $E$ is the gamma-ray photon energy and $n$ is an unknown parameter. When $n=1$, it is shown that a maximum energy-scale of $M_1=9.23\times 10^{19}$ GeV \cite{Vasileiou:2013vra} can be probed, this can be achieved by the lunar array for $\alpha \geq -3$ within 100 hours of integration time. In the case of the extended propagation our results exceed the Lorentz invariance case for all $\alpha \geq -2$ with 10 hours of integration time. For $n=2$, the limits from Lorentz invariance violation are much weaker: that is $M_2=1.3\times 10^{11}$ GeV which is bettered by the lunar array within 1 second. Therefore, in both cases we are able to produce competitive non-observational limits on the energy-scale of quantum gravitational effects. We also note that Lorentz invariance violation limits strongly depend on the model-dependent parameter $n$, whereas our results don't have any such unknown dependence. Additionally, even when $\alpha=-3$ and with only 100 dipoles, we can reach within 2 orders of magnitude of $M_{pl}$ with 1 second of integration time on the lunar array. The only caveat being that these limits require that $\nu_c \geq 80$ kHz.
	
One point worth noting is that the need to align the line of sight with a conversion environment, like M31, implies that there will be strong foregrounds from this galaxy itself. These are not immediately quantifiable at relevant frequencies, indeed the lowest frequency M31 radio flux measurement in the NASA Extra-galactic Database \cite{1993MNRAS.263...25H} is at 151 MHz and produces $0.75$ Jy whereas the conversion signals require sensitivity of $\mathcal{O}(10 \, \mu \mathrm{Jy})$ (note that it is necessary to compare to diffuse fluxes only but the aforementioned is a point source observation). We note that more extended surveys, as reported in \cite{1959MNRAS.119..297B,1961MNRAS.122..479B,1972AJ.....77..637D} indicate far higher fluxes at low frequencies in M31 (around 200 Jy), however, to make these comparable to our background all point sources would first need to be subtracted from consideration. We therefore benchmark against the dominant point source emission, as it is available, and it is unlikely the actual diffuse component will be as bright, as even the diffuse emission in the Coma galaxy cluster radio halo falls below 100 Jy at frequencies below 100 MHz~\cite{Thierbach_2002}. In summary, it is very possible that the signal we predict from graviton-photon conversion would be up to 5 orders of magnitude below the foreground. In addition, we note that there may be other isotropic radio background contributions at low frequencies, as argued in \cite{Dowell:2018mdb} but further understanding of these backgrounds will make their subtraction more feasible. As a point of mitigation we note that \cite{2009NewAR} consider a similar foreground problem from the perspective of observing epoch of reionisation signals and conclude it to be necessary for a lunar array to consist of $10^3$ to $10^5$ elements to observe a non-isotropic signal like the 21 cm power-spectrum. The lower limit of this array size requirement fits conveniently with the number of antennae needed to detect graviton-photon conversion, with a minimum cut-off frequency of 80 kHz. In addition, synchrotron spectra that dominate low-frequency galaxy emissions~\cite{1992ARA&A..30..575C} will have a positive power-law exponent with frequency. Whereas, since it is cutting off, our graviton-photon signal will have a negative slope. This potentially provides a valuable characteristic that can be leveraged to extract the signal despite large foregrounds. This mitigating factor might be cancelled out by free-free absorption modifying the spectral character of the background when viewed through a galactic conversion environment~\cite{1992ARA&A..30..575C}. Additionally, we note that extraction of the signal in the face of the foregrounds and backgrounds remains the largest uncertainty here, as discussed in \cite{2009NewAR}. Thus, this would require precise calibration of the foreground and background fluxes in order to extract the desired signal. We stress, however, that for the foreground and background fluxes any solution to similar challenges in global 21 cm observations can be leveraged. This can be illustrated by the extensive work that has gone into finding methods of extracting the 21 cm global signal in the face of similarly daunting foregrounds~\cite{21cm0,21cm1,21cm2,21cm3,21cm4}. These works have illustrated that robust techniques exist by which low amplitude signals can be extracted in the face of foregrounds several orders of magnitude brighter and should be applicable beyond just the global 21 cm case~\cite{21cm3}. In terms of backgrounds we note that it may be possible to mitigate these via differential observations, as the conversion signal will only be visible when looking at the conversion environment target, whereas other isotropic radio backgrounds will be visible off target as well. This is somewhat similar to the techniques used to measure the Sunyaev-Zel'dovich effect which is typically several orders of magnitude weaker than the cosmic microwave background that it perturbs. However, the effect on potential graviton-photon conversion cannot be estimated without precise knowledge of the conversion environment diffuse foreground and instrumental systematics~\cite{21cm2}.

\begin{figure}[h!]
	\centering
	\includegraphics[width=0.49\textwidth]{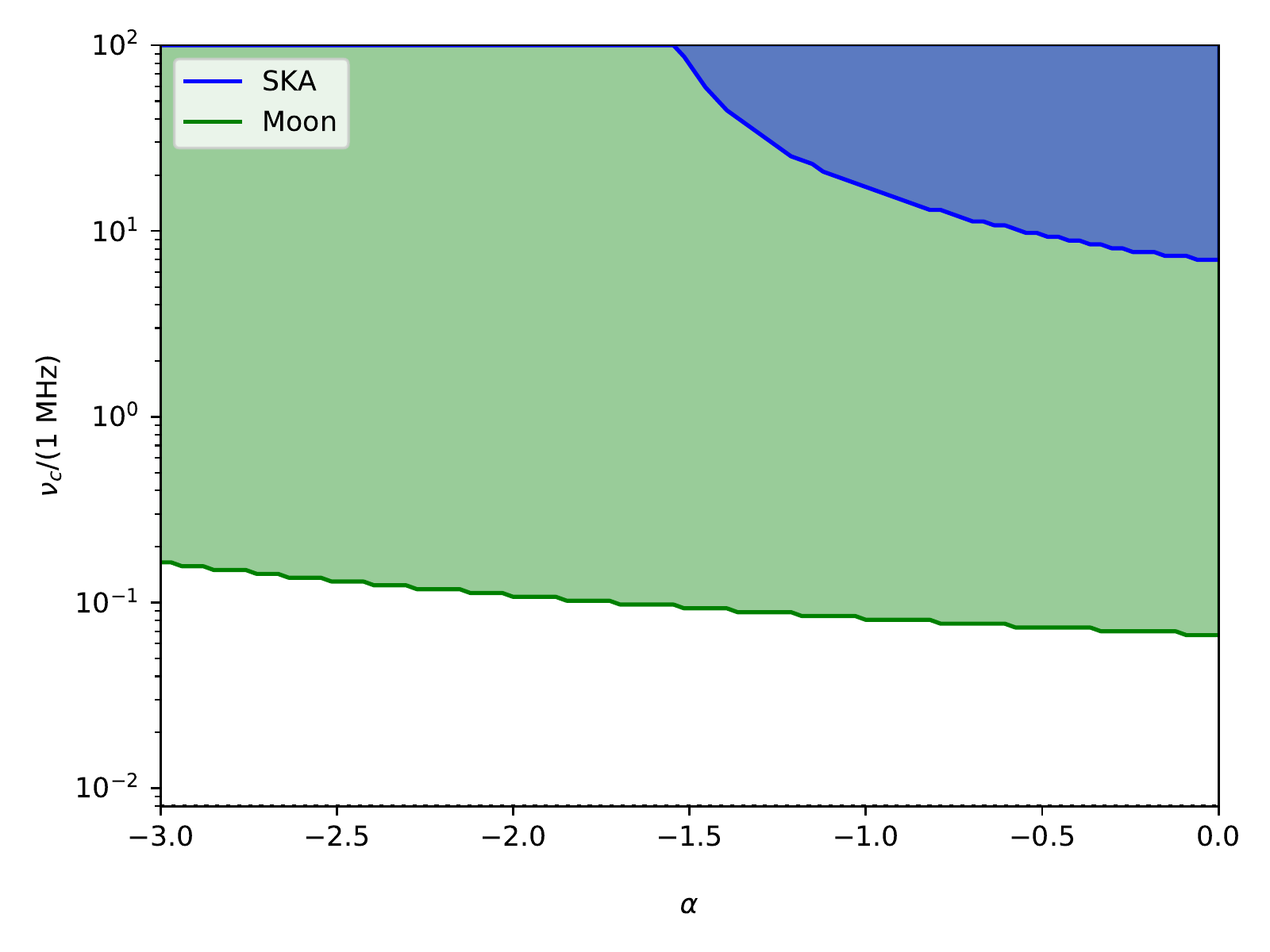}
	\includegraphics[width=0.49\textwidth]{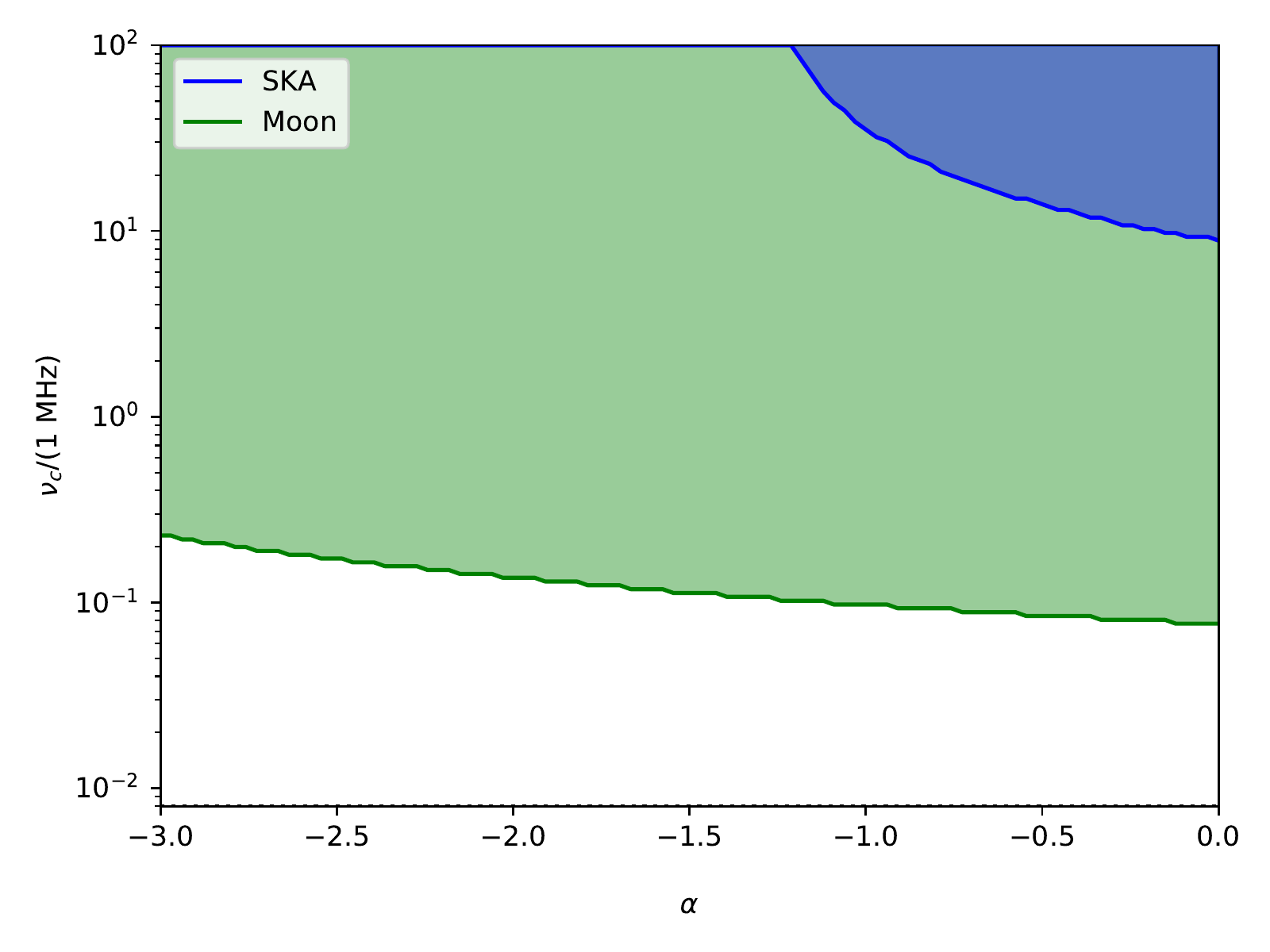}
	
	\includegraphics[width=0.49\textwidth]{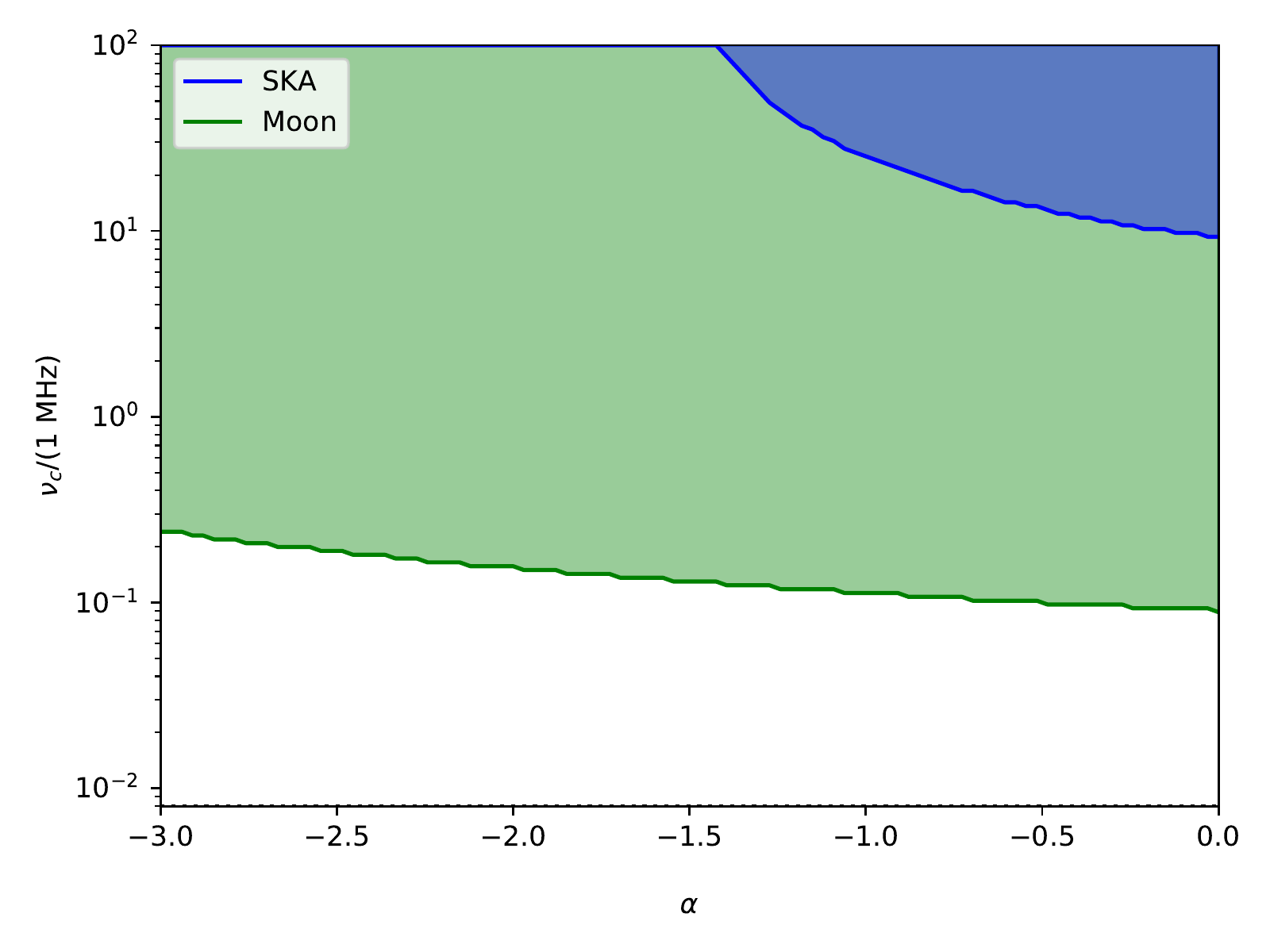}
	\includegraphics[width=0.49\textwidth]{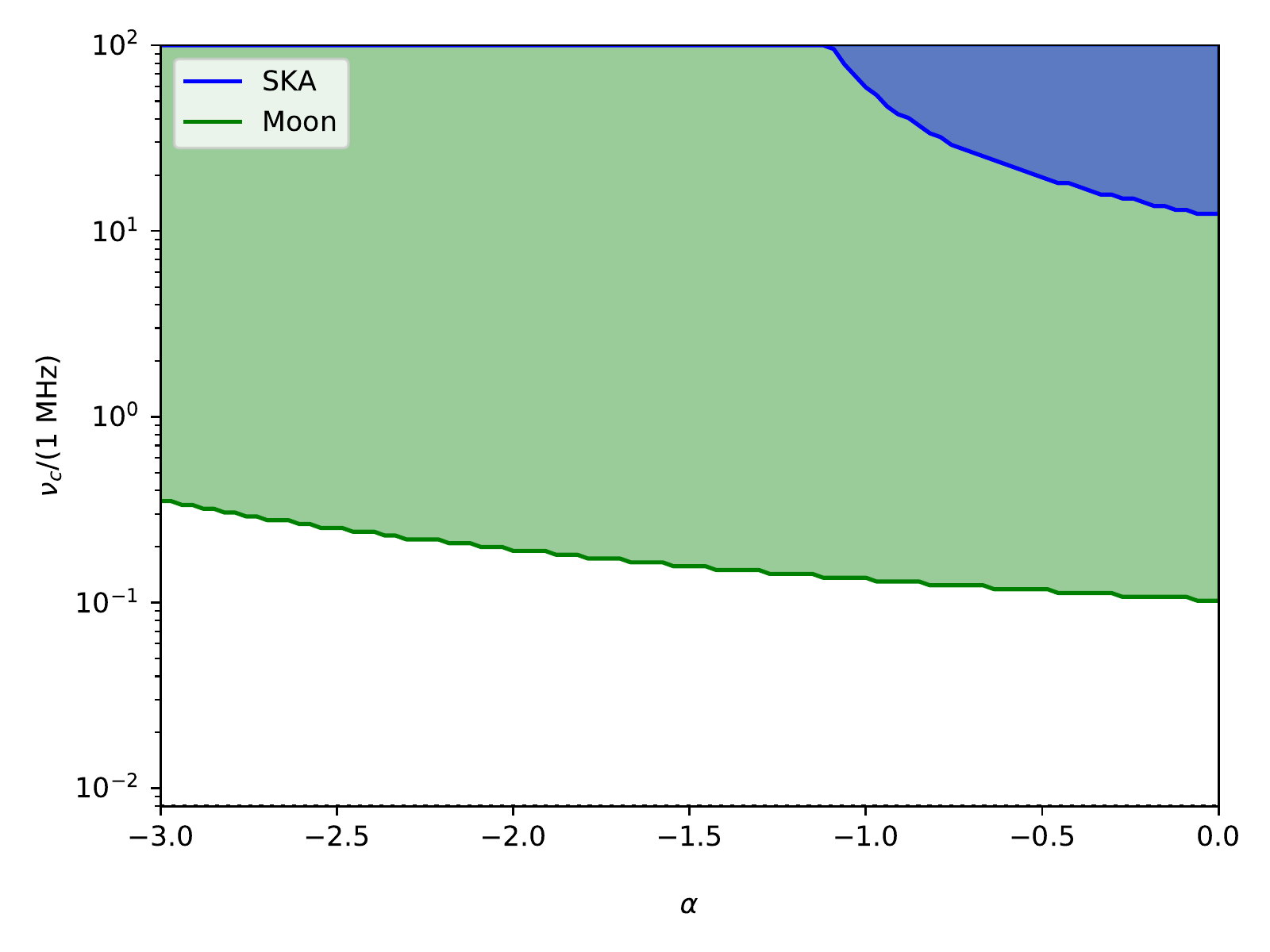}
	\caption{Detectability parameter space for the conversion of gravitons, from an unresolved background, into photons as it varies with conversion environment redshift. All plots assume 100 hours of integration time and 1000 antennae in the lunar array. The left column assumes a conversion fraction like the extended M31 scenario whereas the core case is the right-hand column. Top row: $z = 0.5$. Bottom row: $z = 1.0$.}\label{fig:highz}
\end{figure}

One possible approach to limit the foreground impact is the use of a distant galaxy as a conversion environment. This is because the background gravitational wave flux will be less diminished by the adjustment of the lower limit of the integral in Eq.~(\ref{eq:phi-gw}) than the conversion environment foreground flux will be when moved out to a larger redshift. To illustrate this we present fig.~\ref{fig:highz} which displays how the detectability varies as we adjust the conversion environment redshift. For simplicity we assume that it is possible to find an M31-like galaxy at the given redshift. The results suggest that the minimum cut-off frequency required for detection is mildly sensitive to the conversion environment redshift. However, the flux from the conversion environment galaxy will be reduced by a factor of $~10^7$ from that of M31. This approach could be a promising one but requires the knowledge of the presence of a distant galaxy as well some confidence in the assertion that the conversion fraction would be similar to that of well-studied galaxies in the more local universe.

	\section{\label{con}Conclusion}
	
	In the presented work, we have demonstrated that it is possible to detect low-frequency radio counterparts to gravitational-waves using a lunar dipole antenna array. While this was far more challenging for a single neutron star merger event in the local universe (without a lunar array with at least $10^5$ elements). As long as exponential mode damping does not dominate for frequencies below 80 kHz (an order of magnitude above existing spectra~\cite{Tsang_2019} which show no exponential damping below 8 kHz), we have shown that the detection prospects for an unresolved background from binary neutron star mergers are largely robust to considerations of array size, power-law slopes $> -2$, and integration time. However, the best prospects can be obtained for 1000 hours of integration time and $10^5$ antennae (as recommended for reionisation science in  \cite{2009NewAR}). We have also demonstrated that the galaxy M31 provides a conversion environment in the local universe that allows for some prospect of detection, albeit quite unlikely, by numerically solving the equations of motion for photon-graviton mixing~\cite{PhysRevD.99.044022} and using magnetic field and plasma density information drawn from the literature~\cite{beckm31,beck2015,ruiz-granados2010}. Our modelling of the population of binary neutron stars was taken from \cite{Mapelli_2018} and allowed us to compute the expected background signal with the Planck 2015 cosmological parameters~\cite{planck2015}. Crucially, our results reveal that the detectability of a graviton-photon background from neutron star mergers depends upon whether the power-law behaviour of the gravitational wave spectra~\cite{Kawamura:2003hu,Kawamura:2004ah,Tsang_2019} extends until 80 kHz with a power-law slope $\gtrsim -3$ ($-2$ for $10^3$ antennae and 100 hours observation). If exponential damping dominates below this threshold, or the slope is steeper, then even a lunar array of $10^5$ elements will not detect the resultant background. We note that the 80 kHz is an order of magnitude above the end of the spectrum modelled in \cite{Tsang_2019,Kawamura:2003hu,Kawamura:2004ah} and this would therefore seem to make detection highly improbable. However, it is possible to obtain power-law extensions to gravitational wave spectra from mergers from higher-order harmonic contributions~\cite{Berti:2009kk} and binary mergers can generate signals out to 1 MHz in frequency with the aid of higher-order harmonics~\cite{Martinez:2020cdh}.
	
	An important caveat is the issue of backgrounds and foregrounds. Both of which are expected to be an obstacle due the possibility of strong radio backgrounds~\cite{Dowell:2018mdb} and galaxy foregrounds from the conversion environment itself. Thus, as in the case of epoch of reionisation science discussed in \cite{2009NewAR}, these provide the most significant uncertainties and the success of observations discussed here will depend strongly upon characterisation of these backgrounds/foregrounds. We have shown, however, that it may be possible to mitigate the foreground issue with the use of distant galaxies as conversion environments. We note that the background and foreground issues should be addressable following methods being devised to extract global 21 cm signals~\cite{21cm3} combined with potential differential observation techniques. A second caveat is the uncertainty surrounding the neutron star merger rate, with \cite{Mapelli_2018} placing the rate in the local universe in the range $[20,600]$ Gpc$^{-3}$ yr$^{-1}$ thus proving an order of magnitude uncertainty in the flux from graviton-photon conversion of the unresolved merger background. However, the overlap with the local universe range from LIGO/VIRGO~\cite{Abbott_2019,Abbott_2020} of $[250,2810]$ Gpc$^{-3}$ yr$^{-1}$ suggests that reducing uncertainties could improve the detection prospects presented here. 
	
	We note that several works (such as \cite{PhysRevD.99.044022,Gertsenshtein:1962}) have examined the question of whether photon-graviton conversion could be used to detect such merger events. However, our work has presented novel results by both considering an unresolved gravitational wave background rather than just single events as well as using higher frequency numerical calculations from \cite{Kawamura:2003hu,Kawamura:2004ah,Tsang_2019} and considering phenomenological extrapolations thereof. Crucially, we do not base our conclusions on the details of the extrapolation but rather on the relative independence of the detectability on said details. In particular, our lower limits on the energy scale of interaction assume both a power-law fall-off and an $80$ kHz exponential cut-off. It is worth noting that this requires the power-law extrapolation to hold for an order of magnitude beyond current knowledge of the merger physics. Our use of a parameter space of potential extrapolations allows us to determine the required behaviour in order for a converted signal to become observable. Whether the requirements found for observability might be realised in nature can, of course, be tested by future numerical simulations that extend the frequency range of gravitational wave spectrum.
	
	Vital to this endeavour would be the prospect of placing a telescope on the lunar surface. Landing an array on the moon or placing one is orbit is promising, as can be see by the continuing activity in this field of radio astronomy \cite{bergman2009explorer,2013EPSC....8..279B,BURNS2012433,olfar,KLEINWOLT2012167,dsl}. The results presented here demonstrate that such a lunar instrument may have the potential to probe the realm of quantum gravity at radio frequencies, using this mechanism of mixing between photons and low-mass bosons. It may even be possible to rule out or confirm the existence of the graviton. In particular, a non-observation of a neutron star merger induced radio background will place lower limits on the energy scale of putative quantum gravitational effects, subject to merger rate uncertainties. These lower limits are shown here to be competitive with Lorentz invariance violation experiments, as well as easily reaching within one or two orders of magnitude of the Planck scale. All of this, however, is contingent upon the lack of exponential mode damping prior to 80 kHz and the ability to extract the signal in the presence of highly dominant backgrounds and foregrounds.

	\section{Acknowledgements}
	
	JT and GB honour the memory of Prof. Sergio Colafrancesco, who contributed to this paper, and helped shape our academic lives, before his passing in September 2018. We thank several referees for their comments and criticism that have profoundly shaped this work over a long period of time.
	We also thank Dr Dmitry Prokhorov for his invaluable comments. This work is supported by the South African
	Research Chairs Initiative of the Department of Science and Technology
	and National Research Foundation of South Africa (Grant
	No 77948). J.T. acknowledges support from the DST/NRF
	SKA post-graduate bursary initiative. G.B acknowledges support from a National Research Foundation of South Africa Thuthuka grant no. 117969. This research has made use of the NASA/IPAC Extragalactic Database (NED), which is operated by the Jet Propulsion Laboratory, California Institute of Technology, under contract with the National Aeronautics and Space Administration. This work also made use of the WebPlotDigitizer\footnote{\url{http://automeris.io/WebPlotDigitizer/}}.

	\bibliographystyle{elsarticle-num}
	\bibliography{bib_final}

\begin{thebibliography}{10}
\expandafter\ifx\csname url\endcsname\relax
  \def\url#1{\texttt{#1}}\fi
\expandafter\ifx\csname urlprefix\endcsname\relax\def\urlprefix{URL }\fi
\expandafter\ifx\csname href\endcsname\relax
  \def\href#1#2{#2} \def\path#1{#1}\fi

\bibitem{Schnittman:2014jpa}
J.~D. Schnittman, {Coordinated Observations with Pulsar Timing Arrays and
  ISS-Lobster} (2014).
\newblock \href {http://arxiv.org/abs/1411.3994} {\path{arXiv:1411.3994}}.

\bibitem{Essick:2014wwa}
R.~Essick, et~al., {Localization of short duration gravitational-wave
  transients with the early advanced LIGO and Virgo detectors}, ApJ 800~(2)
  (2015) 81.
\newblock \href {http://arxiv.org/abs/1409.2435} {\path{arXiv:1409.2435}}.

\bibitem{Pannarale:2014rea}
F.~Pannarale, F.~Ohme, {Prospects for joint gravitational-wave and
  electromagnetic observations of neutron-star--black-hole coalescing
  binaries}, ApJ Lett. 791 (2014) 7.
\newblock \href {http://arxiv.org/abs/1406.6057} {\path{arXiv:1406.6057}}.

\bibitem{Williamson:2014wma}
A.~R. Williamson, et~al., {Improved methods for detecting gravitational waves
  associated with short gamma-ray bursts}, Phys. Rev. D. 90~(12) (2014) 122004.
\newblock \href {http://arxiv.org/abs/1410.6042} {\path{arXiv:1410.6042}}.

\bibitem{Kasen:2014toa}
D.~Kasen, R.~Fernandez, B.~Metzger, {Kilonova light curves from the disc wind
  outflows of compact object mergers}, MNRAS 450 (2015) 1777--1786.
\newblock \href {http://arxiv.org/abs/1411.3726} {\path{arXiv:1411.3726}}.

\bibitem{Kamble:2014zya}
A.~Kamble, A.~Soderberg, E.~Berger, A.~Zauderer, S.~Chakraborti, P.~Williams,
  {Radio Supernovae in the Local Universe} (2014).
\newblock \href {http://arxiv.org/abs/1401.1221} {\path{arXiv:1401.1221}}.

\bibitem{Dolgov2017}
A.~{Dolgov}, K.~{Postnov}, {Electromagnetic radiation accompanying
  gravitational waves from black hole binaries}, JCAP 2017~(9) (2017) 018.
\newblock \href {http://arxiv.org/abs/1706.05519} {\path{arXiv:1706.05519}}.

\bibitem{PhysRevD.99.044022}
D.~Ejlli, V.~R. Thandlam,
  \href{https://link.aps.org/doi/10.1103/PhysRevD.99.044022}{Graviton-photon
  mixing}, Phys. Rev. D. 99 (2019) 044022.
\newline\urlprefix\url{https://link.aps.org/doi/10.1103/PhysRevD.99.044022}

\bibitem{TheLIGOScientific:2017qsa}
B.~P. Abbott, et~al., {GW170817: Observation of Gravitational Waves from a
  Binary Neutron Star Inspiral}, Phys. Rev. Lett. 119~(16) (2017) 161101.

\bibitem{GBM:2017lvd}
B.~P. Abbott, et~al., {Multi-messenger Observations of a Binary Neutron Star
  Merger}, ApJ Lett. 848~(2) (2017) 12.
\newblock \href {http://arxiv.org/abs/1710.05833} {\path{arXiv:1710.05833}}.

\bibitem{Alexander:2017aly}
K.~D. Alexander, et~al., {The Electromagnetic Counterpart of the Binary Neutron
  Star Merger LIGO/VIRGO GW170817. VI. Radio Constraints on a Relativistic Jet
  and Predictions for Late-Time Emission from the Kilonova Ejecta}, ApJ 848
  (2017) L21.
\newblock \href {http://arxiv.org/abs/1710.05457} {\path{arXiv:1710.05457}}.

\bibitem{Raffelt:1987im}
G.~Raffelt, L.~Stodolsky, {Mixing of the Photon with Low Mass Particles}, Phys.
  Rev. D. 37 (1988) 1237.

\bibitem{Denisov1978}
V.~I. Denisov, {Interaction of a weak gravitational wave with the field of a
  rotating magnetic dipole}, Sov. Phys. JETP 47 (1978) 209.

\bibitem{Gerlach:1974zz}
U.~H. Gerlach, {Beat Frequency Oscillations near Charged Black Holes and Other
  Electrovacuum Geometries}, Phys. Rev. Lett. 32 (1974) 1023--1025.

\bibitem{Gretarsson_2018}
A.~Gretarsson, P.~Jones, D.~Singleton,
  \href{http://dx.doi.org/10.1142/S0218271818470211}{Gravity’s light in the
  shadow of the moon}, International Journal of Modern Physics D 27~(14) (2018)
  1847021.
\newblock \href {https://doi.org/10.1142/s0218271818470211}
  {\path{doi:10.1142/s0218271818470211}}.
\newline\urlprefix\url{http://dx.doi.org/10.1142/S0218271818470211}

\bibitem{Gertsenshtein:1962}
M.~E. Gertsenshtein, {Wave resonance of light and gravitational waves}, Sov.
  Phys. JETP 14 (1962) 84.

\bibitem{Peccei:1977hh}
R.~D. Peccei, H.~R. Quinn, Cp conservation in the presence of instantons, Phys.
  Rev. Lett. 38 (1977) 1440--1443.

\bibitem{Kaplan:1985dv}
D.~B. Kaplan, {Opening the Axion Window}, Nucl. Phys. B. 260 (1985) 215--226.

\bibitem{Marklund:1999sp}
M.~Marklund, G.~Brodin, P.~K.~S. Dunsby, {Radio wave emissions due to
  gravitational radiation}, ApJ 536 (2000) 875--879.
\newblock \href {http://arxiv.org/abs/astro-ph/9907350}
  {\path{arXiv:astro-ph/9907350}}.

\bibitem{Fargion1995}
D.~Fargion, {Radio Bangs at Kilohertz by SN 1987A: a Test for Graviton-Photon
  Conversion}, Gravitation and Cosmology 1 (1995) 301--310.
\newblock \href {http://arxiv.org/abs/astro-ph/9604047}
  {\path{arXiv:astro-ph/9604047}}.

\bibitem{Mapelli_2018}
M.~Mapelli, N.~Giacobbo, \href{http://dx.doi.org/10.1093/mnras/sty1613}{The
  cosmic merger rate of neutron stars and black holes}, Monthly Notices of the
  Royal Astronomical Society 479~(4) (2018) 4391–4398.
\newblock \href {https://doi.org/10.1093/mnras/sty1613}
  {\path{doi:10.1093/mnras/sty1613}}.
\newline\urlprefix\url{http://dx.doi.org/10.1093/mnras/sty1613}

\bibitem{BURNS2012433}
J.~O. Burns, J.~Lazio, S.~Bale, J.~Bowman, R.~Bradley, C.~Carilli,
  S.~Furlanetto, G.~Harker, A.~Loeb, J.~Pritchard,
  \href{http://www.sciencedirect.com/science/article/pii/S0273117711007472}{Probing
  the first stars and black holes in the early universe with the dark ages
  radio explorer (dare)}, Advances in Space Research 49~(3) (2012) 433 -- 450.
\newblock \href {https://doi.org/https://doi.org/10.1016/j.asr.2011.10.014}
  {\path{doi:https://doi.org/10.1016/j.asr.2011.10.014}}.
\newline\urlprefix\url{http://www.sciencedirect.com/science/article/pii/S0273117711007472}

\bibitem{KLEINWOLT2012167}
M.~K. Wolt], A.~Aminaei, P.~Zarka, J.-R. Schrader, A.-J. Boonstra, H.~Falcke,
  \href{http://www.sciencedirect.com/science/article/pii/S0032063312002796}{Radio
  astronomy with the european lunar lander: Opening up the last unexplored
  frequency regime}, Planetary and Space Science 74~(1) (2012) 167 -- 178,
  scientific Preparations For Lunar Exploration.
\newblock \href {https://doi.org/https://doi.org/10.1016/j.pss.2012.09.004}
  {\path{doi:https://doi.org/10.1016/j.pss.2012.09.004}}.
\newline\urlprefix\url{http://www.sciencedirect.com/science/article/pii/S0032063312002796}

\bibitem{bergman2009explorer}
J.~E.~S. Bergman, R.~J. Blott, A.~B. Forbes, D.~A. Humphreys, D.~W. Robinson,
  C.~Stavrinidis, First explorer -- an innovative low-cost passive
  formation-flying system (2009).
\newblock \href {http://arxiv.org/abs/0911.0991} {\path{arXiv:0911.0991}}.

\bibitem{2013EPSC....8..279B}
R.~J. {Blott}, I.~W.~A. {Baan}, A.~J. {Boonstra}, J.~{Bergman}, D.~{Robinson},
  D.~{Liddle}, N.~{Navarathinam}, S.~{Eves}, C.~{Bridges}, S.~{Gao},
  M.~{Bentum}, A.~{Forbes}, D.~{Humphreys}, C.~G. {Harroch}, {Space-based
  ultra-long wavelength radio observatory (low cost) - SURO-LC}, in: European
  Planetary Science Congress, 2013, pp. EPSC2013--279.

\bibitem{olfar}
S.~Engelen, C.~Verhoeven, M.~Bentum, Olfar, a radio telescope based on nano
  satellites in moon orbit, in: 24th Annual Conference on Small Satellites,
  Utah State University, 2010, pp. 1--7.

\bibitem{Rajan_2015}
R.~T. Rajan, A.-J. Boonstra, M.~Bentum, M.~Klein-Wolt, F.~Belien, M.~Arts,
  N.~Saks, A.-J. van~der Veen,
  \href{http://dx.doi.org/10.1007/s10686-015-9486-6}{Space-based aperture array
  for ultra-long wavelength radio astronomy}, Experimental Astronomy 41~(1-2)
  (2015) 271–306.
\newblock \href {https://doi.org/10.1007/s10686-015-9486-6}
  {\path{doi:10.1007/s10686-015-9486-6}}.
\newline\urlprefix\url{http://dx.doi.org/10.1007/s10686-015-9486-6}

\bibitem{dsl}
A.~Boonstra, M.~Garrett, G.~Kruithof, M.~Wise, A.~{Van Ardenne}, J.~Yan, J.~Wu,
  J.~Zheng, E.~Gill, J.~Guo, M.~Bentum, J.~Girard, X.~Hong, T.~An, H.~Falcke,
  M.~Klein-Wolt, S.~Wu, W.~Chen, L.~Koopmans, H.~Rothkaehl, X.~Chen, M.~Huang,
  L.~Chen, L.~Gurvits, P.~Zarka, B.~Cecconi, H.~{De Haan}, Discovering the sky
  at the longest wavelengths (dsl), in: 2016 IEEE Aerospace Conference, AERO
  2016, Vol. 2016-June, IEEE, United States, 2016.
\newblock \href {https://doi.org/10.1109/AERO.2016.7500678}
  {\path{doi:10.1109/AERO.2016.7500678}}.

\bibitem{dsl1}
L.~{Chen}, M.~{Zhang}, Y.~{Yan}, M.~{Huang}, The concept of space ultra long
  wavelength array, in: 2014 XXXIth URSI General Assembly and Scientific
  Symposium (URSI GASS), 2014, pp. 1--4.

\bibitem{Kawamura:2003hu}
M.~Kawamura, et~al., {General relativistic numerical simulation on coalescing
  binary neutron stars and gauge-invariant gravitational wave extraction}
  (2003).
\newblock \href {http://arxiv.org/abs/astro-ph/0306481}
  {\path{arXiv:astro-ph/0306481}}.

\bibitem{Kawamura:2004ah}
M.~Kawamura, K.~Oohara, {Gauge - invariant gravitational wave extraction from
  coalescing binary neutron stars}, Prog. Theor. Phys. 111 (2004) 589--594.
\newblock \href {http://arxiv.org/abs/astro-ph/0404228}
  {\path{arXiv:astro-ph/0404228}}.

\bibitem{Tsang_2019}
K.~W. Tsang, T.~Dietrich, C.~Van Den~Broeck,
  \href{http://dx.doi.org/10.1103/PhysRevD.100.044047}{Modeling the postmerger
  gravitational wave signal and extracting binary properties from future binary
  neutron star detections}, Physical Review D 100~(4) (Aug 2019).
\newblock \href {https://doi.org/10.1103/physrevd.100.044047}
  {\path{doi:10.1103/physrevd.100.044047}}.
\newline\urlprefix\url{http://dx.doi.org/10.1103/PhysRevD.100.044047}

\bibitem{Vasileiou:2013vra}
V.~Vasileiou, A.~Jacholkowska, F.~Piron, J.~Bolmont, C.~Couturier, J.~Granot,
  F.~W. Stecker, J.~Cohen-Tanugi, F.~Longo, {Constraints on Lorentz Invariance
  Violation from Fermi-Large Area Telescope Observations of Gamma-Ray Bursts},
  Phys. Rev. D. 87~(12) (2013) 122001.
\newblock \href {http://arxiv.org/abs/1305.3463} {\path{arXiv:1305.3463}}.

\bibitem{Dowell:2018mdb}
J.~Dowell, G.~B. Taylor, {The Radio Background Below 100 MHz}, Astrophys. J.
  Lett. 858~(1) (2018) L9.
\newblock \href {http://arxiv.org/abs/1804.08581} {\path{arXiv:1804.08581}},
  \href {https://doi.org/10.3847/2041-8213/aabf86}
  {\path{doi:10.3847/2041-8213/aabf86}}.

\bibitem{2009NewAR}
S.~Jester, H.~Falcke, Science with a lunar low-frequency array: From the dark
  ages of the universe to nearby exoplanets, New Astron. Rev. 53 (2009) 1--26.
\newblock \href {http://arxiv.org/abs/0902.0493} {\path{arXiv:0902.0493}}.

\bibitem{beckm31}
R.~Beck, The magnetic field in m31, A\&A 106~(1) (1982) 121.

\bibitem{ruiz-granados2010}
B.~Ruiz-Granados, J.~A. Rubi{\~{n}}o-Mart{\'{\i}}n, E.~Florido, E.~Battaner,
  \href{https://doi.org/10.1088%2F2041-8205%2F723%2F1%2Fl44}{Magnetic fields
  and the outer rotation curve of m31}, ApJ 723~(1) (2010) L44--L48.
\newblock \href {https://doi.org/10.1088/2041-8205/723/1/l44}
  {\path{doi:10.1088/2041-8205/723/1/l44}}.
\newline\urlprefix\url{https://doi.org/10.1088%2F2041-8205%2F723%2F1%2Fl44}

\bibitem{beck2015}
R.~{Beck}, {Magnetic fields in spiral galaxies}, The Astronomy and Astrophysics
  Review 24 (2015) 4.
\newblock \href {http://arxiv.org/abs/1509.04522} {\path{arXiv:1509.04522}},
  \href {https://doi.org/10.1007/s00159-015-0084-4}
  {\path{doi:10.1007/s00159-015-0084-4}}.

\bibitem{Horns:2012kw}
D.~Horns, et~al., {Hardening of TeV gamma spectrum of AGNs in galaxy clusters
  by conversions of photons into axion-like particles}, Phys. Rev. D. 86 (2012)
  075024.
\newblock \href {http://arxiv.org/abs/1207.0776} {\path{arXiv:1207.0776}}.

\bibitem{Berti:2009kk}
E.~Berti, V.~Cardoso, A.~O. Starinets, {Quasinormal modes of black holes and
  black branes}, Class. Quant. Grav. 26 (2009) 163001.
\newblock \href {http://arxiv.org/abs/0905.2975} {\path{arXiv:0905.2975}},
  \href {https://doi.org/10.1088/0264-9381/26/16/163001}
  {\path{doi:10.1088/0264-9381/26/16/163001}}.

\bibitem{Martinez:2020cdh}
J.~G. Martinez, B.~Kamai, {Searching for MHz gravitational waves from harmonic
  sources}, Class. Quant. Grav. 37~(20) (2020) 205006.
\newblock \href {http://arxiv.org/abs/2010.06118} {\path{arXiv:2010.06118}},
  \href {https://doi.org/10.1088/1361-6382/aba669}
  {\path{doi:10.1088/1361-6382/aba669}}.

\bibitem{Abbott_2019}
B.~Abbott, R.~Abbott, T.~Abbott, S.~Abraham, F.~Acernese, K.~Ackley, C.~Adams,
  R.~Adhikari, V.~Adya, C.~Affeldt, et~al.,
  \href{http://dx.doi.org/10.1103/PhysRevX.9.031040}{Gwtc-1: A
  gravitational-wave transient catalog of compact binary mergers observed by
  ligo and virgo during the first and second observing runs}, Physical Review X
  9~(3) (Sep 2019).
\newblock \href {https://doi.org/10.1103/physrevx.9.031040}
  {\path{doi:10.1103/physrevx.9.031040}}.
\newline\urlprefix\url{http://dx.doi.org/10.1103/PhysRevX.9.031040}

\bibitem{Abbott_2020}
B.~P. Abbott, R.~Abbott, T.~D. Abbott, S.~Abraham, F.~Acernese, K.~Ackley,
  C.~Adams, R.~X. Adhikari, V.~B. Adya, C.~Affeldt, et~al.,
  \href{http://dx.doi.org/10.3847/2041-8213/ab75f5}{Gw190425: Observation of a
  compact binary coalescence with total mass $\sim$ 3.4 m$_\odot$}, The
  Astrophysical Journal 892~(1) (2020) L3.
\newblock \href {https://doi.org/10.3847/2041-8213/ab75f5}
  {\path{doi:10.3847/2041-8213/ab75f5}}.
\newline\urlprefix\url{http://dx.doi.org/10.3847/2041-8213/ab75f5}

\bibitem{planck2015}
P.~A.~R. Ade, N.~Aghanim, M.~Arnaud, M.~Ashdown, J.~Aumont, C.~Baccigalupi,
  A.~J. Banday, R.~B. Barreiro, J.~G. Bartlett, et~al.,
  \href{http://dx.doi.org/10.1051/0004-6361/201525830}{Planck2015 results},
  Astronomy \& Astrophysics 594 (2016) A13.
\newblock \href {https://doi.org/10.1051/0004-6361/201525830}
  {\path{doi:10.1051/0004-6361/201525830}}.
\newline\urlprefix\url{http://dx.doi.org/10.1051/0004-6361/201525830}

\bibitem{Chen_2018}
L.~Chen, A.~Aminaei, L.~I. Gurvits, M.~K. Wolt, H.~R. Pourshaghaghi, Y.~Yan,
  H.~Falcke, \href{http://dx.doi.org/10.1007/s10686-018-9576-3}{Antenna design
  and implementation for the future space ultra-long wavelength radio
  telescope}, Experimental Astronomy 45~(2) (2018) 231–253.
\newblock \href {https://doi.org/10.1007/s10686-018-9576-3}
  {\path{doi:10.1007/s10686-018-9576-3}}.
\newline\urlprefix\url{http://dx.doi.org/10.1007/s10686-018-9576-3}

\bibitem{refId0}
{M. P. van Haarlem}, et~al.,
  \href{https://doi.org/10.1051/0004-6361/201220873}{Lofar: The low-frequency
  array}, A\&A 556 (2013) A2.
\newblock \href {https://doi.org/10.1051/0004-6361/201220873}
  {\path{doi:10.1051/0004-6361/201220873}}.
\newline\urlprefix\url{https://doi.org/10.1051/0004-6361/201220873}

\bibitem{5109716}
S.~W. {Ellingson}, T.~E. {Clarke}, A.~{Cohen}, J.~{Craig}, N.~E. {Kassim},
  Y.~{Pihlstrom}, L.~J. {Rickard}, G.~B. {Taylor}, The long wavelength array,
  Proceedings of the IEEE 97~(8) (2009) 1421--1430.

\bibitem{Tingay_2013}
S.~J. Tingay, R.~Goeke, J.~D. Bowman, D.~Emrich, S.~M. Ord, D.~A. Mitchell,
  M.~F. Morales, T.~Booler, B.~Crosse, R.~B. Wayth, et~al.,
  \href{http://dx.doi.org/10.1017/pasa.2012.007}{The murchison widefield array:
  The square kilometre array precursor at low radio frequencies}, Publications
  of the Astronomical Society of Australia 30 (2013).
\newblock \href {https://doi.org/10.1017/pasa.2012.007}
  {\path{doi:10.1017/pasa.2012.007}}.
\newline\urlprefix\url{http://dx.doi.org/10.1017/pasa.2012.007}

\bibitem{report_1997}
P.~Bely, R.~Laurance, S.~Volonte, R.~Ambrosini, A.~Ardenne, C.~Barrow,
  J.~Bougeret, J.~Marcaide, G.~Woan, Very low frequency array on the lunar far
  side, ESA Report SCI(97)2 (1997).

\bibitem{2009newarray}
S.~Jester, H.~Falcke, Science with a lunar low-frequency array: From the dark
  ages of the universe to nearby exoplanets, New Astron. Rev. 53 (2009) 1.
\newblock \href {http://arxiv.org/abs/0902.0493} {\path{arXiv:0902.0493}}.

\bibitem{ska2012}
P.~Dewdney, others., SKA baseline design document (2012).

\bibitem{1742-6596-653-1-012139}
E.~A. Lisin, et~al., Lunar dusty plasma: A result of interaction of the solar
  wind flux and ultraviolet radiation with the lunar surface, J. of Phys.
  Conference Series 653~(1) (2015) 012139.

\bibitem{earthIonosphere}
D.~Anderson, T.~Fuller-Rowell,
  \href{http://solar-center.stanford.edu/SID/science/Ionosphere.pdf}{The
  ionosphere}, Space Environment Topics (1999).
\newline\urlprefix\url{http://solar-center.stanford.edu/SID/science/Ionosphere.pdf}

\bibitem{DeBreuck:2000zk}
C.~De~Breuck, W.~van Breugel, H.~J.~A. Rottgering, G.~Miley, {A Sample of 669
  ultrasteep spectrum radio sources to find high redshift radio galaxies}, A\&A
  143 (2000) 303--333.
\newblock \href {http://arxiv.org/abs/astro-ph/0002297}
  {\path{arXiv:astro-ph/0002297}}.

\bibitem{1993MNRAS.263...25H}
S.~E.~G. {Hales}, J.~E. {Baldwin}, P.~J. {Warner}, {The 6C survey of radio
  sources - VI.}, MNRAS 263 (1993) 25--30.
\newblock \href {https://doi.org/10.1093/mnras/263.1.25}
  {\path{doi:10.1093/mnras/263.1.25}}.

\bibitem{1959MNRAS.119..297B}
R.~H. {Brown}, C.~{Hazard}, {The radio emission from normal galaxies, I.
  Observations of M31 and M33 at 158 Mc/s and 237 Mc/s}, MNRAS 119 (1959) 297.
\newblock \href {https://doi.org/10.1093/mnras/119.3.297}
  {\path{doi:10.1093/mnras/119.3.297}}.

\bibitem{1961MNRAS.122..479B}
R.~H. {Brown}, C.~{Hazard}, {The radio emission from normal galaxies, II},
  MNRAS 122 (1961) 479.
\newblock \href {https://doi.org/10.1093/mnras/122.6.479}
  {\path{doi:10.1093/mnras/122.6.479}}.

\bibitem{1972AJ.....77..637D}
J.~M. {Durdin}, Y.~{Terzian}, {Low Frequency Radio Observations of the
  Andromeda Galaxy}, Astronomical Journal 77 (1972) 637.
\newblock \href {https://doi.org/10.1086/111329} {\path{doi:10.1086/111329}}.

\bibitem{Thierbach_2002}
M.~Thierbach, U.~Klein, R.~Wielebinski,
  \href{http://dx.doi.org/10.1051/0004-6361:20021474}{The diffuse radio
  emission from the coma cluster at 2.675 ghz and 4.85 ghz}, Astronomy \&
  Astrophysics 397~(1) (2002) 53–61.
\newblock \href {https://doi.org/10.1051/0004-6361:20021474}
  {\path{doi:10.1051/0004-6361:20021474}}.
\newline\urlprefix\url{http://dx.doi.org/10.1051/0004-6361:20021474}

\bibitem{1992ARA&A..30..575C}
J.~J. {Condon}, {Radio emission from normal galaxies.}, Annual Review of
  Astronomy and Astrophysics 30 (1992) 575--611.
\newblock \href {https://doi.org/10.1146/annurev.aa.30.090192.003043}
  {\path{doi:10.1146/annurev.aa.30.090192.003043}}.

\bibitem{21cm0}
G.~J.~A. Harker, J.~R. Pritchard, J.~O. Burns, J.~D. Bowman,
  \href{https://doi.org/10.1111/j.1365-2966.2011.19766.x}{{An MCMC approach to
  extracting the global 21-cm signal during the cosmic dawn from sky-averaged
  radio observations}}, Monthly Notices of the Royal Astronomical Society
  419~(2) (2011) 1070--1084.
\newblock \href
  {http://arxiv.org/abs/https://academic.oup.com/mnras/article-pdf/419/2/1070/3107911/mnras0419-1070.pdf}
  {\path{arXiv:https://academic.oup.com/mnras/article-pdf/419/2/1070/3107911/mnras0419-1070.pdf}},
  \href {https://doi.org/10.1111/j.1365-2966.2011.19766.x}
  {\path{doi:10.1111/j.1365-2966.2011.19766.x}}.
\newline\urlprefix\url{https://doi.org/10.1111/j.1365-2966.2011.19766.x}

\bibitem{21cm1}
A.~Liu, J.~R. Pritchard, M.~Tegmark, A.~Loeb,
  \href{http://dx.doi.org/10.1103/PhysRevD.87.043002}{Global 21 cm signal
  experiments: A designer’s guide}, Physical Review D 87~(4) (Feb 2013).
\newblock \href {https://doi.org/10.1103/physrevd.87.043002}
  {\path{doi:10.1103/physrevd.87.043002}}.
\newline\urlprefix\url{http://dx.doi.org/10.1103/PhysRevD.87.043002}

\bibitem{21cm2}
S.~Singh, R.~Subrahmanyan, N.~U. Shankar, A.~Raghunathan,
  \href{http://dx.doi.org/10.1088/0004-637X/815/2/88}{On the detection of
  global 21-cm signal from reionization using interferometers}, The
  Astrophysical Journal 815~(2) (2015) 88.
\newblock \href {https://doi.org/10.1088/0004-637x/815/2/88}
  {\path{doi:10.1088/0004-637x/815/2/88}}.
\newline\urlprefix\url{http://dx.doi.org/10.1088/0004-637X/815/2/88}

\bibitem{21cm3}
K.~Tauscher, D.~Rapetti, J.~O. Burns, E.~Switzer,
  \href{https://doi.org/10.3847%2F1538-4357%2Faaa41f}{Global 21 cm signal
  extraction from foreground and instrumental effects. i. pattern recognition
  framework for separation using training sets}, The Astrophysical Journal
  853~(2) (2018) 187.
\newblock \href {https://doi.org/10.3847/1538-4357/aaa41f}
  {\path{doi:10.3847/1538-4357/aaa41f}}.
\newline\urlprefix\url{https://doi.org/10.3847%2F1538-4357%2Faaa41f}

\bibitem{21cm4}
K.~Tauscher, \href{https://baas.aas.org/pub/aas236-321p02-tauscher}{Robust
  extraction of the cosmological global 21-cm signal from foreground and
  instrumental systematic effects}, Bulletin of the AAS 52~(3),
  https://baas.aas.org/pub/aas236-321p02-tauscher (6 2020).
\newline\urlprefix\url{https://baas.aas.org/pub/aas236-321p02-tauscher}

\end{thebibliography}

\end{document}